\documentclass[times,final]{elsarticle}

 \RequirePackage{geometry}
\def\snm#1{\textcolor{red}{#1}}  
\usepackage{framed,multirow}
\usepackage[utf8]{inputenc}
\usepackage[T1]{fontenc}
\usepackage{amssymb,amsmath}
\usepackage{latexsym}
\usepackage{algorithmicx,algorithm,algpseudocode}
\usepackage{url}
\usepackage{siunitx}
\usepackage{xcolor}
\definecolor{newcolor}{rgb}{.8,.349,.1}

\usepackage[center]{subfigure}
\usepackage{epstopdf}
\newcommand{\ud}{\, \mathrm{d}}	
\newcommand{\abs}[1]{\left\lvert #1\right\rvert}

\renewcommand{\i}{\mathrm{i}}
\newcommand{\e}{\mathrm{e}}
\newcommand{\nuv}{\boldsymbol{\nu}}
\newcommand{\Lv}{\mathbf{L}}
\newcommand{\xv}{\mathbf{x}}
\newcommand{\vv}{\mathbf{v}}
\newcommand{\nv}{\mathbf{n}}
\newcommand{\Fv}{\mathbf{F}}
\newcommand{\Kv}{\mathbf{K}}
\newcommand{\Mv}{\mathbf{M}}
\newcommand{\kv}{\mathbf{k}}
\newcommand{\kappav}{\boldsymbol\kappa}	
\newcommand{\n}{\mathbf{n}}

\definecolor{bluefoam}{rgb}{0.68, 0.68, 0.82}
\newcommand{\LegendDashdot}{$^{\line(1,0){6}} \, ^{\line(1,0){.5}} \, ^{\line(1,0){6}} \;$} 
\newcommand{\LegendLine}{$^{\line(1,0){15}} \;$} 
\newcommand{\LegendDotted}{$^. \; ^. \; ^. \;$} 

\newcommand{\LegendDashed}{$^{\line(1,0){4}} \hspace{5pt} ^{\line(1,0){4}} \;$} 

\begin{document}


\begin{frontmatter}

\title{A high order continuation method to locate exceptional points and to compute Puiseux series with applications to acoustic waveguides}

\author[1]{Benoit \snm{Nennig}\corref{cor1}}
\cortext[cor1]{Corresponding author}
\ead{benoit.nennig@supmeca.fr}
\author[2]{Emmanuel \snm{Perrey-Debain}}
\ead{Emmanuel.Perrey-Debain@utc.fr}

\address[1]{Institut sup\'erieur de m\'ecanique de Paris (SUPMECA), Laboratoire Quartz EA 7393,\\
3 rue Fernand Hainaut, 93407 Saint-Ouen, France.}
\address[2]{Sorbonne universités, Université de Technologie de Compiègne,\\
Laboratoire Roberval, UMR CNRS 7337, CS 60319, 60203 Compiègne cedex, France.}

\date{\today}
\begin{abstract}
A numerical algorithm is proposed to explore in a systematic way the trajectories of the eigenvalues of non-Hermitian matrices in the parametric space and exploit this in order to find the locations of defective eigenvalues in the complex plane. These non-Hermitian degeneracies also called exceptional points (EP) have raised considerable attention in the scientific community as these can have a great impact in a variety of physical problems. 

The method requires the computation of successive derivatives of two selected eigenvalues with respect to the parameter so that, after recombination, regular functions can be  constructed. This algebraic manipulation permits the localization of exceptional points (EP), using standard root-finding algorithms and the computation of the associated Puiseux series up to an arbitrary order.
This representation, which is associated with the
topological structure of Riemann surfaces allows  to efficiently approximate the selected pair in a certain neighbourhood of the EP. 

Practical applications dealing with guided acoustic waves propagating in straight ducts with absorbing walls and in periodic guiding structures are given to illustrate the versatility of the proposed method and its ability to handle large size matrices arising from 
finite element discretization techniques.
The fact that EPs are associated with optimal dissipative treatments in the sense that they should provide best modal attenuation is also discussed.
\end{abstract}

\begin{keyword}
Exceptional point\sep defective eigenvalue \sep Puiseux series \sep parametric eigenvalue problem
\sep duct acoustics \sep bordered matrix  
\end{keyword}

\end{frontmatter}




\section{Introduction}

In many configurations, the description of the propagation of waves in lossless uniform or periodic guiding structures
can be accomplished via discretization of a wave equation which generally yield eigenvalue problems involving real symmetric (or sometimes Hermitian) matrices. By nature, these matrices are always diagonalizable and associated eigenvectors which are related to the mode shapes in the waveguide, satisfy standard orthogonality properties. This is no longer true if, for instance, the system is open, the media is lossy or if the walls of the waveguide are absorbing or contains active components.
These scenarios are associated with non-Hermitian matrices, often complex symmetric, which means that eigenvectors are not orthogonal any more. 
In some applications, as this will be illustrated in the present paper, these matrices can be regarded as function of one (or more) 
complex-valued parameter and the system may become defective for specific values of the latter.

Finding defective eigenvalues associated with parametric non-Hermitian eigenvalue problems have raised considerable attention in the scientific community as these can have a great impact in a variety of physical problems \cite{Berry:2004,Heiss:1990,Heiss:2012}. These non-Hermitian degeneracies are associated with the existence of exceptional points~\cite{Kato:1980} (EPs) in the complex plane. At the EP, eigenvalues and associated eigenvectors coalesce at a branch point singularity in the parametric space and in the vicinity of the EP, perturbation analysis show that eigenvalues take the form of a Puiseux series \cite{Kato:1980} i.e. a series in fractional power of the parameter. Although not diagonalizable, spectral analysis can be conducted using the Jordan normal form \cite{Trefethen:2005} for the defective matrix.

Parametric eigenvalue problems and EPs have been the subject of intensive research in the domain of classical mechanics \cite{Seyranian:2005,Triantafyllou:1991,SeyranianBook:2003}, acoustics \cite{Tester:1973,Shenderov:2000,Bi:2015,Bloch3D:2017,Achilleos:2017}, optics \cite{Feng:2014}, 
quantum mechanics \cite{Moiseyev:2011,Uzdin:2010,Jarlebring:2011,Cartarius:2009,Cartarius:2016} or in  phase transition
in PT-symmetry systems \cite{Bender:2018,Achilleos:2017} and metamaterials \cite{Bloch3D:2017,Achilleos:2017}, to quote but a few.
In the domain of structural dynamics, the location of EP also have consequences on sensitivity analysis in \cite{Adhikari:2007,CSMA:2019} as well as on veering phenomena \cite{Dieci:2014}.

 The existence of EPs has also created connexion between several mathematical fields such as multiple roots of non-linear equations, Jordan form of defective matrices, bi-orthogonality relations \cite{Lawrie:1999,Moiseyev:2011,Redon:2011}, algebraic geometry and Puiseux series, bifurcation theory and branch tracking or branch point in the Riemann surface. All these facets of the EP are presented in a recent paper of Hanson et al. \cite{Hanson:2019}.



In the scientific litterature, different methods have  been proposed and developed in order to track exceptional points and find the behaviour of the eigenvalues in its vicinity.
One approach is based on a closed loop in the parameter space \cite{Hernandez:2005,Cartarius:2009,Uzdin:2010}. In the case of an EP of order 2 for instance, the square root behavior of two coalescing eigenvalues is exploited. This permits to detect an EP if two eigenvalues permute and do not return to their initial value after a cycle. This property is a direct consequences of the branch point topological structure. This approach although very robust requires a large number of function evaluations.  
The branch point singularity is also exploited in \cite{Lefebvre:2010,Landau:2015} using multipoint Padé approximants.
Here, the distribution of poles and zeros of the Padé approximated function lies along the branch cut where the function is multivalued \cite{Stahl:1997,Aptekarev:2015} which allows to give a first estimate of the EP in the complex plane.
An alternative is to exploit the analyticity of the matrix determinant. Akinola \cite{Akinola:2014} extends the implicit determinant method developed in \cite{Spence:2005} to compute efficiently the determinant of an arbitrary matrix
thanks to the bordered matrix and Cramer's method. The approach is extended to calculate the derivatives of the determinant allowing to recover the first terms of the Puiseux series expansion and to compute the Jordan blocks of standard eigenvalue problem. This approach is scalable and can be used on sparse matrices.

Other methods use the explicit form of the Puiseux series expansion of the eigenvalues in the neighborhood of the EP.
Uzdin et al. \cite{Uzdin:2010} propose a 3 points method in order to provide an estimate  of the first terms of the Puiseux series. At each point an eigenvalue problem is solved and a new estimation of the EP position is obtained by solving a small linear system.
The approach has been extended by Cartarius et al.~\cite{Cartarius:2009,Cartarius:2016} using  a 8 points iterative scheme working either for 2 real or one complex parameter. 
Another strategy is to exploit the normal Jordan form of the defective matrix. 
In Mailybaev's approach \cite{Mailybaev:2006}, which is based on the smoothness of the multiple eigenvalues surfaces in parameter space, Jordan blocks
are computed iteratively from the a priori knowledge of the Jordan structure. The method, which is developed for the analysis of standard eigenvalue problems with multiparameter dependency, allows estimating the Wilkinson's distance (see also \cite{Akinola:2014b}) and applies to defective eigenvalues of high order.
The Jordan form is also used by Welters \cite{Welters:2011} which utilizes closed form expressions in order to obtain high order terms of the Puiseux series.
In situations where matrices are nearly defective, eigensolvers are likely to deliver poor results. In order to mitigate this difficulty, Hernandez \cite{Hernandez:2017} proposes a scalable algorithm to find the normal Jordan form once the location of the EP has been found with sufficient accuracy. The method determines the subspace spanned by the nearly defective eigenvectors using a modified invert iteration method or using direct sparse solvers.
The  parametric eigenvalue problem  can also be recast into a sequence of multiple eigenvalue problems as shown in \cite{Jarlebring:2011,Muhivc:2014}. The approach allows to compute EPs and their associated defective eigenvalue in the case of second-order degeneracy \cite{Jarlebring:2011}. The method is suitable for arbitrary matrices but limited to linear dependency with respect to the parameter. It requires the treatment of matrices of size $\mathcal{N}^2\times \mathcal{N}^2$ for an initial problem of size $\mathcal{N}\times \mathcal{N}$ making the method very prohibitive for large size matrices.
Finally, Zsuzsanna, \cite{Zsuzsanna:2018} proposes to locate EP by solving constrained minimization problems using Newton-type iterative methods.

In this paper, a direction of particular interest  to us is to investigate the existence of  EPs for the analysis
 of sound wave propagation in guiding structures. The idea is not new  and 
the occurrence of multiple eigenvalues in sound-absorbing ducts has been the subject of intensive research since the 70's \cite{Tester:1973}.
It was observed that this particular scenario which occurs for discrete values of the wall admittance corresponds to the coalescence 
of two (or more) successive acoustic duct modes and usually provides optimal sound attenuation rates.
This feature has been exploited in many works for liner design, especially for turbofan ducts (see references in \cite{Bi:2015}).
The connection between EP and sound attenuation in ducts has recently been re-visited in \cite{Bi:2015,Bloch3D:2017}. 
Most of the works just quoted deal with relatively standard duct geometries, circular, rectangular or annular with locally reacting boundary conditions
at the wall. This permits a mathematical description of guided modes using closed-form expressions and the localisation of EPs can be achieved 
via standard root-finding algorithms.

In order to deal with more complex configurations, the use of discretization methods such as the finite element method
is unavoidable. In this case, techniques must be devised in order to deal efficiently with  matrices, usually of relatively large size and sparse, stemming from the discretization of the wave equation.
Our aim is to propose a general method applicable to a large class of problems, which should extend beyond the scope of the  study of guided waves, in order to  identify the existence of exceptional points in the complex plane and to
explore the trajectories of the eigenvalues in its neighborhood.
The  method developed here is based on the calculation of high order derivatives of a selected pair of eigenvalues, corresponding to 
two successive modes susceptible to coalesce,  using the bordered matrix approach \cite{Andrew:1993}. Following similar ideas presented in \cite{Dieci:2014,Uzdin:2010}, the loss  of regularity of the eigenvalues near EP is circumvented thanks to two functions, which by construction, are regular in the neighbourhood of the EP where the two eigenvalues must coalesce. This allows to locate an EP, with great precision if necessary, as well as to compute the  associated Puiseux series up to an arbitrary order. The main ingredients of the method are explained in Sections 2 and 3. In section 4, two configurations are presented in order to show the numerical efficiency of the method both in terms of computational complexity and accuracy.

\section{General statement on eigenvalue pertubation}\label{sec:Perturb}

We are interested in the reconstruction of a selected pair of eigenvalues which must be regarded as function of a single complex-valued parameter $\nu$
 associated with the matrix eigenvalue problem of size $\mathcal{N}\times \mathcal{N}$,

\begin{equation}\label{eq:PVP}
\Lv (\lambda(\nu), \nu) \xv (\nu) =\mathbf{0},
\end{equation}
where $\lambda(\nu)$ is an eigenvalue and $\xv(\nu)\neq \mathbf{0}$ is the right eigenvector. Eq.~\eqref{eq:PVP} is voluntarily written in its general form
though in most applications, as some illustrative examples will be presented in section~\ref{sec:Examples}, the matrix $\Lv$ corresponds to generalized linear or quadratic eigenvalue problems. In the specific context of waveguide modal decomposition, eigenvalues represent the mode axial wavenumbers. In principle, the dependence with respect to the parameter $\nu$ does not need to be explicit though it is generally known thus allowing 
analytical calculation of derivatives.
Following results of \cite[Theorem 2.1]{Andrew:1993}, if is assumed that $\Lv$ is an analytic function of both $\nu$ and $\lambda$ in the neighborhood of $\nu_0$, 
then an eigenvalue $\lambda$ and its associated eigenvector are also analytic in a neighbourhood of $\nu_0$.
This means that 
\begin{equation}
	\lambda(\nu) = \sum_{n=0}^\infty \frac{\lambda^{(n)}(\nu_0)}{n!}  \left(\nu- \nu_0\right)^n,
\end{equation}
where upperscript $^{(n)}$ denotes the $n$\textsuperscript{th} derivative with respect to $\nu$. For this result to hold, it is also assumed that $\lambda(\nu_0)$ must be a simple eigenvalue.

Things are dramaticaly different if the eigenvalue is defective which can happen whenever the parameter $\nu$ corresponds 
to an exceptional point, call it $\nu^*$. Assuming that the associated eigenvalue $\lambda^*$ is of algebraic multiplicity 2,  
the behavior of the two branches of solutions in the vicinity of $\nu^*$ is given by a convergent Puiseux series \cite{Kato:1980}
\begin{subequations}\label{eq:Puiseux}
\begin{align}
\lambda_+ &= a_0 + a_1\left(\nu - \nu^*\right)^\frac{1}{2} + \sum_{k=2}^\infty a_k  \left(\nu - \nu^*\right)^{\frac{k}{2}},\\
\lambda_- &= a_0 - a_1\left(\nu - \nu^*\right)^\frac{1}{2} + \sum_{k=2}^\infty a_k \left( -1 \right)^k \left(\nu - \nu^*\right)^{\frac{k}{2}},
\end{align}
\end{subequations}
where $a_0=\lambda^*$. As stated in \cite{Seyranian:2005,Welters:2011}, as long as $a_1\neq 0$ the two branches coalesce at a branch point singularity  when $\nu$ tends to $\nu^*$. 
In the neighborhood of the EP, eigenvalues exhibit an infinite sensitivity, since the derivatives of the eigenvalue are singular.
In some cases, it can happen that $a_1=0$ which means that all fractional terms in Eq.~\eqref{eq:Puiseux} vanish \cite{Seyranian:2005,Welters:2011} and the matrix is not defective anymore. 
Though the analysis is limited here to two eigenvalues, Puiseux series can be generalized to describe the  coalescence of $m$ eigenvalues as shown in~\cite{Kato:1980} for the case of standard eigenvalue problems, and it can be shown that
eigenvectors also admit a Puiseux series~\cite{Kato:1980,Baumgartel:1985}.
By nature, there is a single eigenvector at the EP and the matrix is not diagonalizable. Nevertheless, it is possible to compute its Jordan canonical form,  see for instance \cite{Akinola:2014,Welters:2011,Mailybaev:2006} for standard eigenvalue problems and also an extension of the method \cite{Guttel:2017} for more general problems. 
%
The Jordan form is also exploited  by Welters \cite{Welters:2011} who proposes a numerical recursive algorithm to compute all Puiseux series terms from matrix derivatives with respect to the parameter.

\section{Localization of exceptional points  and eigenvalue reconstruction}

\subsection{Analytic auxiliary function}
The proposed algorithm exploits the knowledge of high order derivatives of two selected eigenvalues denoted $\lambda_+$ and  $\lambda_-$ and calculated at an arbitrary value $\nu_0$. 
In order to circumvent the branch point singularity in Eq.~\eqref{eq:Puiseux}, two auxiliary functions are defined:
\begin{align}
g(\nu) &= \lambda_+ + \lambda_-,\label{eq:g}\\
h(\nu) &= \left(\lambda_+ - \lambda_- \right)^2\label{eq:h}.
\end{align}
By construction, these functions are holomorphic in the vicinity of $\nu^*$ where eigenvalues coalesce, as this was already mentioned in \cite{Hernandez:2005,Cartarius:2009,Uzdin:2010,Dieci:2014} and in \cite[p. 66]{Kato:1980}. 
The regularity of these two functions can easily be checked by injecting the local behavior given in Eq.~\eqref{eq:Puiseux} in Eqs.~\eqref{eq:g} and  \eqref{eq:h}. 
Note that for linear eigenvalue problems function $h$ can be shown to be related to the discriminant of the characteristic polynomial \cite[remark 3.23]{Dieci:2014} and thus benefit from the same regularity properties as the matrix coefficients.
Function $g(\nu)$ involve even term of the Puiseux series whereas $h(\nu)$ contains odd terms associated to fractional powers and
has been defined to ensure that $h(\nu^*)=0$.

The derivative of $g$ and $h$ at $\nu_0$ can be directly obtained from the derivative of the selected eigenvalues. Applying Liebnitz' rule for product derivation yields
\begin{subequations}
\begin{align}
g^{(n)}(\nu_0) &= \lambda_+^{(n)}(\nu_0) + \lambda_-^{(n)}(\nu_0) ,\label{eq:dg}\\
h^{(n)}(\nu_0) &= \sum_{k=0}^{\lfloor \tfrac{n}{2}\rfloor} \binom{n}{k} \, \left(2-\delta_{\tfrac{n}{2}k}\right)  \left(  \lambda_+^{(n-k)} (\nu_0)\lambda_+^{(k)} (\nu_0)+  \lambda_-^{(n-k)}(\nu_0)\lambda_-^{(k)}(\nu_0) \right) -2 \sum_{k=0}^n \binom{n}{k}\, \lambda_+^{(n-k)}(\nu_0)\lambda_-^{(k)}(\nu_0) , \label{eq:dh}
\end{align}
\end{subequations}
Here, symbol $\lfloor\, \rfloor$ is the floor function and  $\delta_{\tfrac{n}{2}k}$ is the Kronecker symbol generalized to rational numbers and is equal to zero whenever $n$ is odd. 
These notations have been introduced in order to take into account symmetries in the formulation as this can be  used in order to optimize the calculation.
These functions can be approximated, at least locally, via their truncated Taylor series
\begin{subequations}
\begin{align}
\mathcal{T}_g(\nu) &= \sum_{n=0}^N b_n \left(\nu- \nu_0\right)^n,\quad \text{where } b_n=\frac{g^{(n)}(\nu_0)}{n!} ,\label{eq:gTaylor}\\
\mathcal{T}_h(\nu)&= \sum_{n=0}^N c_n \left(\nu - \nu_0\right)^n,\quad \text{where } c_n=\frac{h^{(n)}(\nu_0)}{n!}. \label{eq:hTaylor}
\end{align}
\end{subequations}
In practice, derivatives of the selected pair of eigenvalues can be recursively obtained by solving the bordered matrix~\cite{Andrew:1993} 
of size $\mathcal{N}+1\times \mathcal{N}+1$:
\begin{equation}\label{eq:Bordered}
\begin{bmatrix}
\Lv & \partial_\lambda \Lv \xv \\
\vv^t & 0
\end{bmatrix}
\begin{pmatrix}
\xv^{(n)}\\ \lambda^{(n)}
\end{pmatrix}
= \begin{pmatrix}
\Fv_n\\ 0
\end{pmatrix},
\end{equation}
where the right hand side (RHS) vector $\Fv_n$ contains terms arising from previous order derivatives. Details of this approach are briefly reminded in~\ref{sec:Deriv0}.

\subsection{EP localisation and computation of Puiseux series coefficients}
One of the key ideas of the present approach is to suppose that the EP where the selected eigenvalues must coalesce,
is located in the disc of convergence of the infinite Taylor series of $h$. In this case, it is expected that one of the roots of 
$\mathcal{T}_h$ should provide a good estimate of $\nu^*$. 
Here, the computation of the roots is achieved by calculating the eigenvalues 
of the companion matrix. 
Only few of these roots correspond to the genuine roots of $h$, \textit{ie.} the EPs. The other are spurious roots. A practical way to discriminate between genuine and spurious roots is postponed to the results of secs.~\ref{sec:resultsZ} and \ref{sec:BlochResults}.

Once an approximation of $\nu^*$ is found with sufficient accuracy and confidence, the computation of the Puiseux series can be done in two steps. The first one consists in rewriting the truncated Taylor series of $g$ in the conventional polynomial form

\begin{equation}
\mathcal{T}_g = \sum_{n=0}^N b_n \sum_{k=0}^n \binom{n}{k} (-\nu_0)^{n-k}\nu^k, 
\end{equation}
which, by construction, should be identical to the following alternative form stemming from the truncated Puiseux series \eqref{eq:Puiseux}
\begin{equation}
g \approx \sum_{n=0}^{N} 2 a_{2n} (\nu-\nu^*)^n = \sum_{n=0}^{N} 2 a_{2n} \sum_{k=0}^n \binom{n}{k} (-\nu^*)^{n-k}\nu^k.
\end{equation}
Equating each power of $\nu$ allows to produce a set of equations which can be written in the compact form 
\begin{equation}\label{eq:ae}
2 \mathbf{P}(\nu^*) \mathbf{a}_{\mathrm{e}} = \mathbf{P}(\nu_0) \mathbf{b},
\end{equation}
that relates the unknown even coefficient $a_{2n}$ of the Puiseux series gathered in the vector $\mathbf{a}_{\mathrm{e}} = (a_0,a_2,a_4,\ldots)^t$ to those of the Taylor series collected in $\mathbf{b}$. 
Note that large values of $\abs{\nu^*}$ might  affect the condition number of the matrix $\mathbf{P}(\nu^*) $ and this can be simply avoided  by  making the change of variable \( \nu'=\nu-\nu^* \).  This yields 
\begin{equation}
 2 \mathbf{a}_e = \mathbf{P}(\nu_0-\nu^*) \mathbf{b},
\end{equation}
which also does not necessitate a matrix inversion.
Each term of the upper triangular matrix  $\mathbf{P}(\xi)$ is given by
\begin{equation}
\left(\mathbf{P}(\xi) \right)_{kn}= \left\{
\begin{aligned}
&\binom{n}{k} \, (-\xi)^{n-k}, &&\text{if } n\ge k \\
&0, &&\text{otherwise}.
\end{aligned}\right.
\end{equation}
Note the matrix is of size $2\lfloor \frac{N}{2}\rfloor $ which corresponds to the highest order involved.
This procedure can be interpreted as finding the connecting coefficients between the two Taylor series expansion.
The odd terms of the Puiseux series are obtained in a similar manner using the truncated Taylor series of $h$.
It is convenient, for this purpose, to operate the change of variable $\nu'=\nu-\nu^*$ so we can write
\begin{equation}
\mathcal{T}_h = \sum_{n=0}^N c_n (\nu'-(\nu_0-\nu^*))^n = \sum_{n=0}^N c_n \sum_{k=0}^n \binom{n}{k} (\nu^*-\nu_0)^{n-k}\nu'^k
= \sum_{k=0}^{N}(\mathbf{P}(\nu_0-\nu^*) \mathbf{c})_k \, \nu'^k.
\end{equation}
After substitution of the Puiseux series, we also have
\begin{equation}\label{eq:hPuiseux}
h(\nu) = 4 \left(  a_1{\nu'}^\frac{1}{2}+  a_3{\nu'}^\frac{3}{2} + \dots \right)^2
= 4 \sum_{k=0,1,2,\ldots} \nu'^k \sum_{n=1,3,\dots,2k-1}
 a_n \, a_{2k-n} .
\end{equation}
Equating each power of $\nu'$ gives explicitly the first coefficient for $k=1$,
\begin{equation}
a_1 = \pm \frac{1}{2} \sqrt{(\mathbf{P}(\nu_0-\nu^*) \mathbf{c})_1},
\end{equation}
where the sign refers to one of two branches of the Puiseux series given in Eq.~\eqref{eq:Puiseux}. 
The others coefficients are obtained iteratively as
\begin{equation}\label{eq:ao}
a_{2k-1} =  \frac{1}{8 a_1} \left\{ (\mathbf{P}(\nu_0-\nu^*) \mathbf{c})_k\, - 4\sum_{\substack{n=3,5,\ldots,2k-3\\ n \ge 2k-3}}
	a_n \, a_{2k-n} \right\}, \quad \text{for } k=2,3,\ldots.
\end{equation}

\subsection{Implementation details}\label{sec:Comput}

Eigenvalue computations are performed away from EP whereby matrices are non-defective so standard numerical algorithms, suitable for the treatment of relatively large size matrices, can be used. For the example described in Section~\ref{sec:Bloch3D} PETSc \cite{Petsc} (Portable, Extensible Toolkit for Scientific Computation)  and its Python bindings Petsc4py \cite{Dalcin:2011} are used for parallel computing, data treatment and for the interface management with both the linear solver Mumps \cite{Mumps} and the eigensolvers from SLEPc/ Slepc4py \cite{SLEPc,Dalcin:2011} (Scalable Library for Eigenvalue Problem Computations). In particular, the first eigenvalues having the smallest magnitude for the quadratic eigenvalue problem of Section~\ref{sec:Bloch3D} are obtained via the Two-level Orthogonal Arnoldi (TOAR) method. This algorithm is dedicated to polynomial eigenvalue problem which is suitable for general Hermitian as well as non-Hermitian matrices.

Derivatives of eigenvalues are computed following Andrew's bordered matrix approach which is briefly reminded in the Appendix.
For each selected eigenvalue, linear direct solvers are used to factorize the bordered matrix \eqref{eq:Bordered} and multiple right-hand side vectors $\Fv_n$ are treated by forward and backward substitution. These linear solvers are also used for the eigenvalue computation using shift and invert spectral transformations.

We may note that in order to handle general parametric dependency, it is mandatory to have access to the derivatives of the operator and in the present 
work, derivatives are computed analytically.
In situations where the operator derivatives are not available, which is the case when using with black-box FEM solvers, numerical differentiation using  finite differences can be used, especially if the operator is a linear function of the parameter. Such approach has been used for instance in~\cite{Ghienne:2017}.

The proposed approach has been implemented in a dedicated open source Python library called EasterEig, available at
\url{https://github.com/nennigb/EasterEig}. The example of sec. \ref{sec:impedance} is also provided in the ``examples'' folder.

\section{Examples related to duct acoustics}\label{sec:Examples}

Acoustic treatments are often used to attenuate sound propagation in guiding structures generally encountered in HVAC, exhaust devices, and aircraft engines. In  many applications, the walls of the waveguide are acoustically treated with absorbing materials, generally porous,  and it is usual for the analysis to distinguish between two classes of liners depending whether the latter is locally or non-locally reacting.
In both cases, the propagation of sound waves is best described in terms of duct acoustic modes and,
as this was discussed earlier in the introduction, the existence of EP is known to be connected with very strong modal attenuation.

The two following examples of growing complexity are chosen in order to illustrate the accuracy, the robustness as well as to identify some limitations of the method. A very classical example is first investigated in Section~\ref{sec:impedance} for which exact solutions to the problem
can be found in the literature. The more complex configuration treated in  Section~\ref{sec:Bloch3D}, which is not amenable to analytical treatment, is investigated in order to assess other technical aspects of the method such as computational complexity and execution time.

\subsection{Acoustic duct with locally reacting liner}\label{sec:impedance}

\subsubsection{Finite element model}

We consider a  two-dimensional acoustic waveguide of infinite length of height $h_d$. In the duct, the acoustic pressure satisfies the Helmholtz equation ($\e^{-\i \omega t}$ convention is adopted here)
\begin{equation}\label{eq:Helmholtz}
\Delta p + k_a^2 p =0,
\end{equation}
where $k_a=\omega/c_a$ is the wavenumber, $c_a$ is the sound speed and $\omega$ is the angular frequency.
At the lined wall $y=0$, the liner is assumed to be locally reacting which implies that 
the pressure must satisfy the impedance boundary condition
\begin{equation}
Z = \frac{p}{\vv \cdot \nv}.
\end{equation}
where the vector $\nv$ is the outgoing unit normal to the wall surface and the acoustic velocity $\vv$  is proportional to the pressure gradient via the Euler equation: $\i \omega \rho_a \vv =  \nabla p  $ where $\rho_a$ is the fluid density.
The real part of the impedance $Z$ is related to dissipation mechanisms whereas the imaginary part accounts for all restoring effects, inertia and stiffness, of the absorbing  material. 
On the upper wall $y=h_d$, rigid wall condition is applied and $\vv\cdot\nv=0$. 
Using invariance along the waveguide $x$-axis, the modal analysis is performed by assuming that the pressure field can be written in the separable form
$p = \phi(y) \e^{\i \beta x }$. Here, function $\phi$ is the mode shape and $\beta$ the axial wavenumber.  
The weak formulation associated with the Helmoltz equation gives
\begin{equation}
- \int_0^{h_d} \partial_y \psi \cdot \partial_y \phi \ud y + (k_a^2 - \beta^2)\int_0^{h_d}  \psi \phi \ud y
+ \frac{\i \omega\rho_a}{Z} \psi(0)  \phi(0)=0,
\end{equation}
where $\psi$ stands for the test function. Once the variational formulation is discretized with linear Lagrangian finite element, we obtain a generalized eigenvalue problem of the form \eqref{eq:PVP}. To be consistent with our notation, we put $\lambda=\beta^2$ and $\nu=Z$ as the complex-valued parameter of the problem so we have 
\begin{equation}\label{eq:PVPz}
\Lv (\lambda(\nu), \nu) \xv(\nu)  =  (-\Kv + (k_a^2-\lambda) \Mv + \i \omega\rho_a \, \nu^{-1} \boldsymbol\Gamma) \xv(\nu) =\mathbf{0}.
\end{equation}
Here, vector $\xv$ contains the finite element nodal values of the acoustic pressure,  the matrix $\boldsymbol\Gamma$ stems from the impedance boundary condition and $\Kv$ and $\Mv$ are the standard stiffness and mass matrices respectively.  
We can easily calculate that 
\begin{equation}
\partial_\lambda \Lv=-\Mv \quad \mathrm{and} \quad \Lv^{(k)}=  \i \omega\rho_a \, (-1)^k k! \nu^{-k-1} \, \boldsymbol\Gamma
\end{equation}
so the right-hand side of  Eq. \eqref{eq:Bordered} can be obtained explicitely as 
\begin{equation}\label{eq:deriv_z}
 \Fv_n =- \i \omega\rho_a \boldsymbol\Gamma \sum_{k=1}^{n} \binom{n}{k} (-1)^k k! \nu^{-k-1} \xv^{(n-k)}  
+ \Mv \sum_{k=1}^{n-1}\binom{n}{k} \lambda^{(k)} \xv^{(n-k)},
\end{equation}
where, it is reminded that matrix $\boldsymbol\Gamma$ contains zero in all entries excepted one term which greatly simplifies the matrix-vector operation in  Eq.~\eqref{eq:deriv_z}.

\subsubsection{Results on accuracy}\label{sec:resultsZ}

In order to validate the accuracy of the proposed method, we refer to the analytical solution due to Tester \cite{Tester:1973}. 
Transverse modes associated with EPs have the general form $\phi_j(y) =\cos (\alpha_j (y-h_d))$ where $\alpha_j$ must obey 
the double roots condition
\begin{equation}\label{eq:ZepAna}
\sin {(2 \alpha_j h_d)} + (2 \alpha_j h_d) =0,
\end{equation}
where the correspondence $j \leftrightarrow \alpha_j$ is given by ordering
the imaginary part in  ascending order.
 Once the roots are found with standard root-finding methods (see~\cite{Nennig:2010} for instance), wall impedance values are obtained by the formula
\begin{equation}\label{eq:ZepAna2}
\nu^*_{j \rightarrow j+1}= -\frac{\i k_a h_d}{(\alpha_j h_d) \tan (\alpha_j h_d)}\rho_a c_a.
\end{equation}
The notation $j \rightarrow j+1$, which is proposed here in order to illustrate and facilitate the analysis of  results given in the next section, means that the EP corresponds to the coalescence of the eigenvalue pair ($\lambda_j$,$\lambda_{j+1}$).
Note that, solutions of Eq. \eqref{eq:ZepAna} admit the asymptotic approximation \cite{Tester:1973} $2 \alpha_j h_d \approx( 4j+3)\pi/2 \pm \i \ln((4j+3)\pi)$
so zero is an accumulation point for the EP series $\nu^*_{j \rightarrow j+1}$. This is reflected by the fact that the 
operator $\Lv$ is singular at $\nu=0$ and this could have been simply avoided if instead, the admittance $Z^{-1}$ was taken as a parameter but to illustrate the robustness of the method it is preferable for the exercise to consider $Z$ as a parameter.

Numerical computations are performed for a duct with unit width, discretized with 100 linear Lagrange finite elements. The physical properties are set to $c_a=340$ \si{m.s^{-1}}, $\rho_a=1.2$ \si{kg.m^{-3}} and $f=200$ \si{Hz}. 
In the following examples, the arbitrary value  $\nu_0$ is chosen in the neighborhood of the EP such that $\nu_0 = \nu^* + \abs{\nu}^* \Delta \e^{\i \pi/4}$, here $\Delta= {\abs{\nu_0-\nu^*}}/{\abs{\nu^*}}$ stands for the relative distance between the two points.
Once the truncated Taylor series $\mathcal{T}_h$ has been constructed, an approximate value of an EP must 
be found among the $N$ roots of the $N$\textsuperscript{th} order polynomial $\mathcal{T}_h$. In order to discriminate the genuine root, i.e. the EP, with 
spurious  roots, we can  refer to the work of \cite{Christiansen:2006} who showed that spurious roots  tend to be aligned on a circle  centered about $\nu_0$ and which radius provides an upper bound for the radius of convergence of the infinite Taylor expansion.
Spurious roots, call them $\zeta_n$ with $n=1,\ldots,N'$ and $N' \leq N$, tend to be evenly distributed around the circle and it is convenient for the analysis to define 
the following indicator, which gives a crude but reasonable value for the radius of convergence of the series,
\begin{equation}\label{eq:rootradius}
\rho^{\mathrm{roots}}_N(\mathcal{T}_{h}) = \frac{1}{N'}\sum_{n=1}^{N'} \abs{\zeta_n-\nu_0}.
\end{equation}

\begin{figure}
  \begin{center}
	\subfigure[Roots of $\mathcal{T}_h(\lambda_1,\lambda_2)$, $\nu^*=445.803 + 357.211\i$,\,\,  \mbox{$\rho^\mathrm{roots}(\mathcal{T}_h)\approx 530 $}.]{\includegraphics[width=0.45\textwidth]{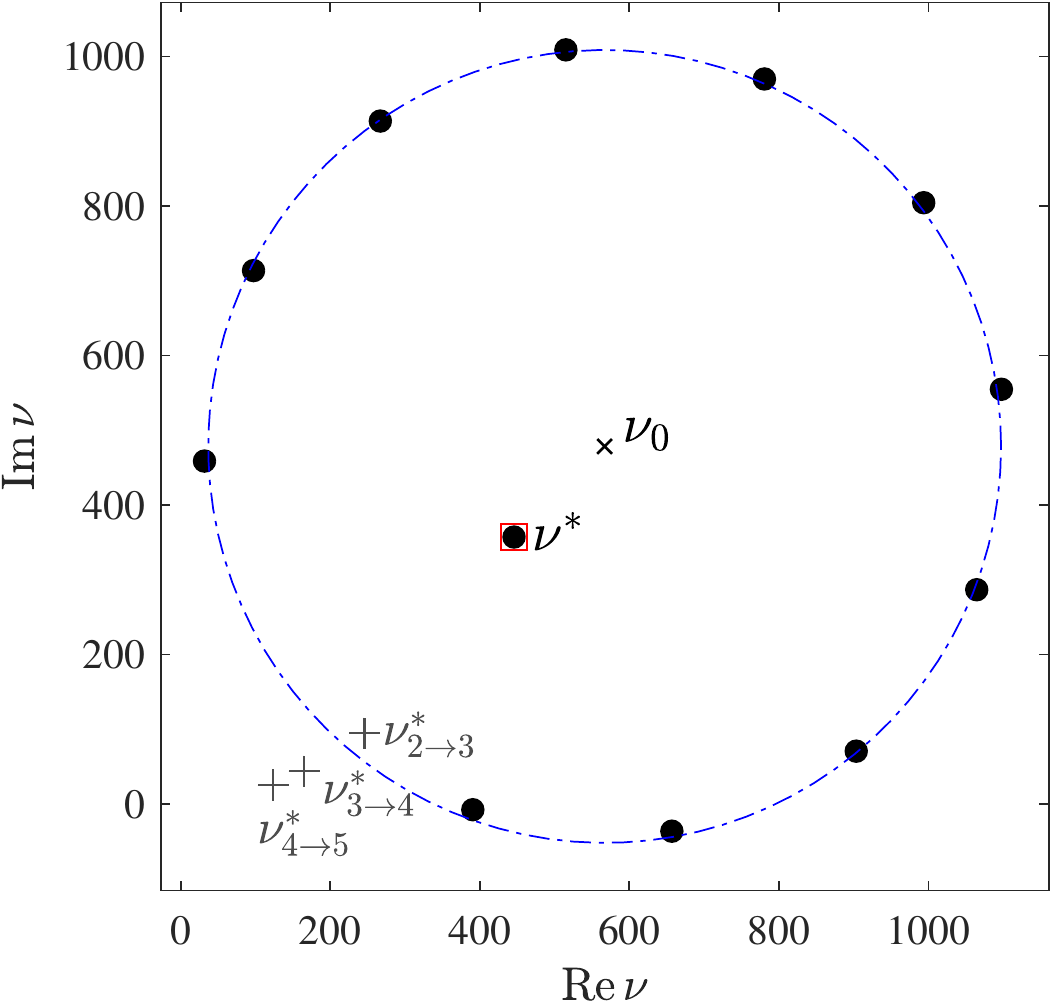}}\hfill
	\subfigure[Roots of $\mathcal{T}_h(\lambda_2,\lambda_3)$, $\nu^*=246.057 + 94.9156\i$,\,\,  \mbox{$\rho^\mathrm{roots}(\mathcal{T}_h)\approx 218 $}.]{\includegraphics[width=0.45\textwidth]{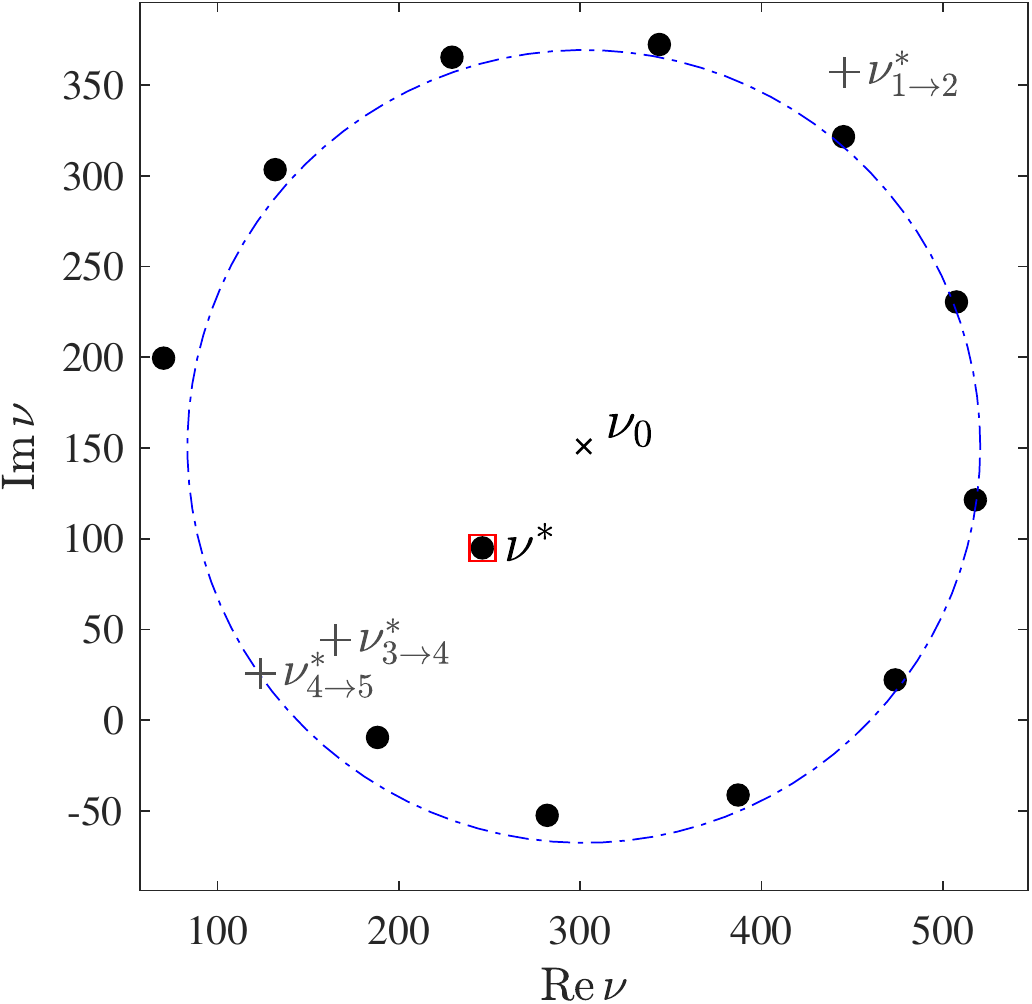}}\\
	\subfigure[Roots of $\mathcal{T}_h(\lambda_3,\lambda_4)$, $\nu^*=165.134 + 44.1474\i$,\,\, \mbox{$\rho^\mathrm{roots}(\mathcal{T}_h)\approx 77 $}.]{\includegraphics[width=0.45\textwidth]{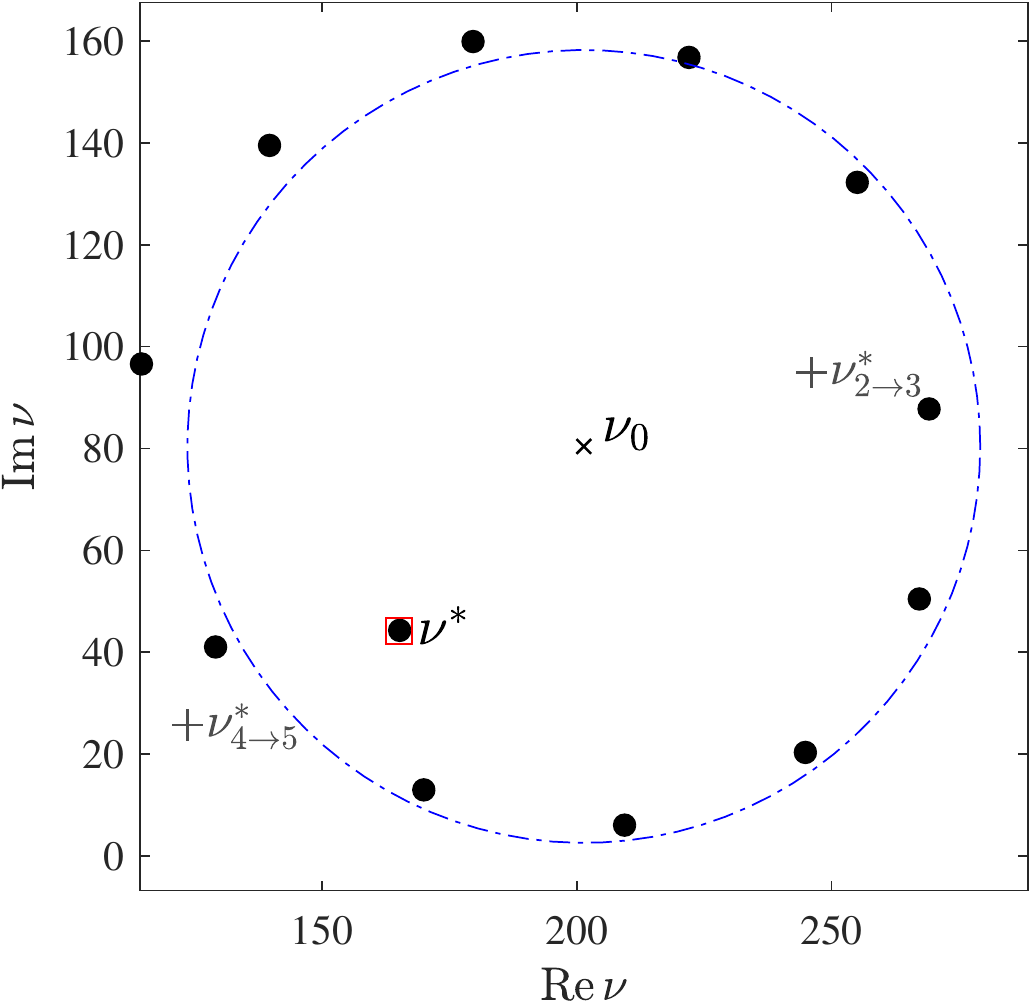}}\hfill
	\subfigure[Roots of $\mathcal{T}_h(\lambda_4,\lambda_5)$, $\nu^*=123.660 +  25.7198\i$,\,\, \mbox{$\rho^\mathrm{roots}(\mathcal{T}_h)\approx 28 $}.]{\includegraphics[width=0.45\textwidth]{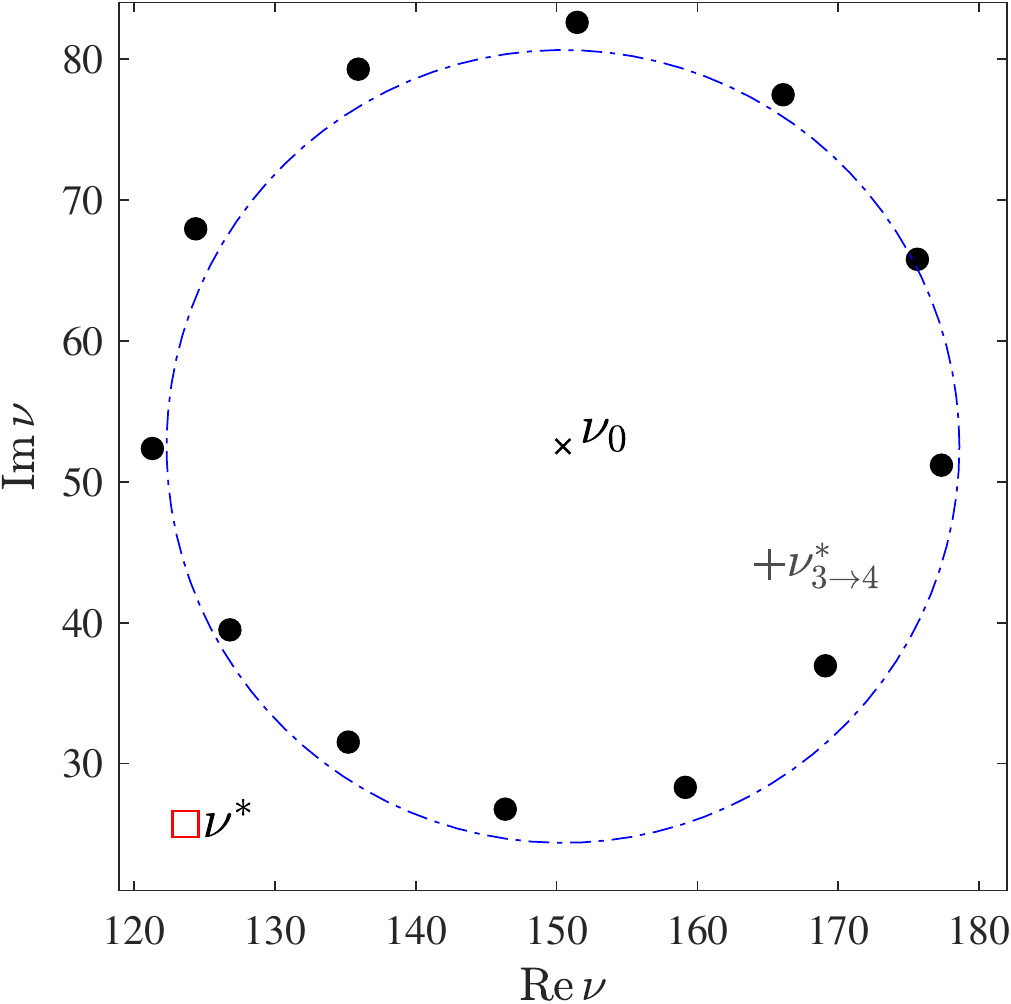}}
   \caption{Location of the complex roots $\zeta_n$ ($\bullet$) of the Taylor expansion $\mathcal{T}_h(\nu)$ with $N= 12$ and $\Delta=0.3$. The circle centered on $\nu_0$ has a radius given by the estimate Eq.~\eqref{eq:rootradius}. The marker $\textcolor{red}{\square}$ indicates the location of the  exceptional point. Symbols \textcolor[gray]{0.4}{$+$} refer to the nearest EP corresponding to the coalescence of another pair of eigenvalues.} 
   \label{fig:ZplaneRoots}
  \end{center}
\end{figure}

In Fig.~\ref{fig:ZplaneRoots}a-d) are reported the location of the roots of $\mathcal{T}_{h}$ in the complex plane for 4 different situations 
corresponding with the first 4 EPs, given in Eq. \eqref{eq:ZepAna2}. In order to distinguish the different scenarios, we call 
$\mathcal{T}_h(\lambda_j,\lambda_{j+1})$ the truncated series related to the coalescence of the pair of eigenvalues $\lambda_j$ and $\lambda_{j+1}$.
In all case, we took $\Delta=0.3$ with $N=12$.
It can be observed that the radius of convergence of the series is limited by the existence of another EP involving one of the eigenvalue from the pair with another eigenvalue, for instance $\nu^*_{3 \rightarrow 4}$ and  $\nu^*_{4 \rightarrow 5}$ corresponding to the pair $(\lambda_3,\, \lambda_4)$ and $(\lambda_4,\, \lambda_5)$ respectively. 
In  Figs.~\ref{fig:ZplaneRoots}a-c), a genuine root is clearly identified  inside the circle and this corresponds to the existence of an EP. However in the last example shown in Fig.~\ref{fig:ZplaneRoots}d) the EP is outside the radius of convergence of $\mathcal{T}_h$ and cannot be found directly
and in this case another initial value for $\nu_0$ has to be provided. 
The fact that the radius of convergence of $h$ is diminishing as $j$ increases is a direct consequence of the behaviour of the EP sequence which converges towards zero.

\begin{figure}
  \begin{center}                                   
 \subfigure[]{\includegraphics[width=0.48\textwidth]{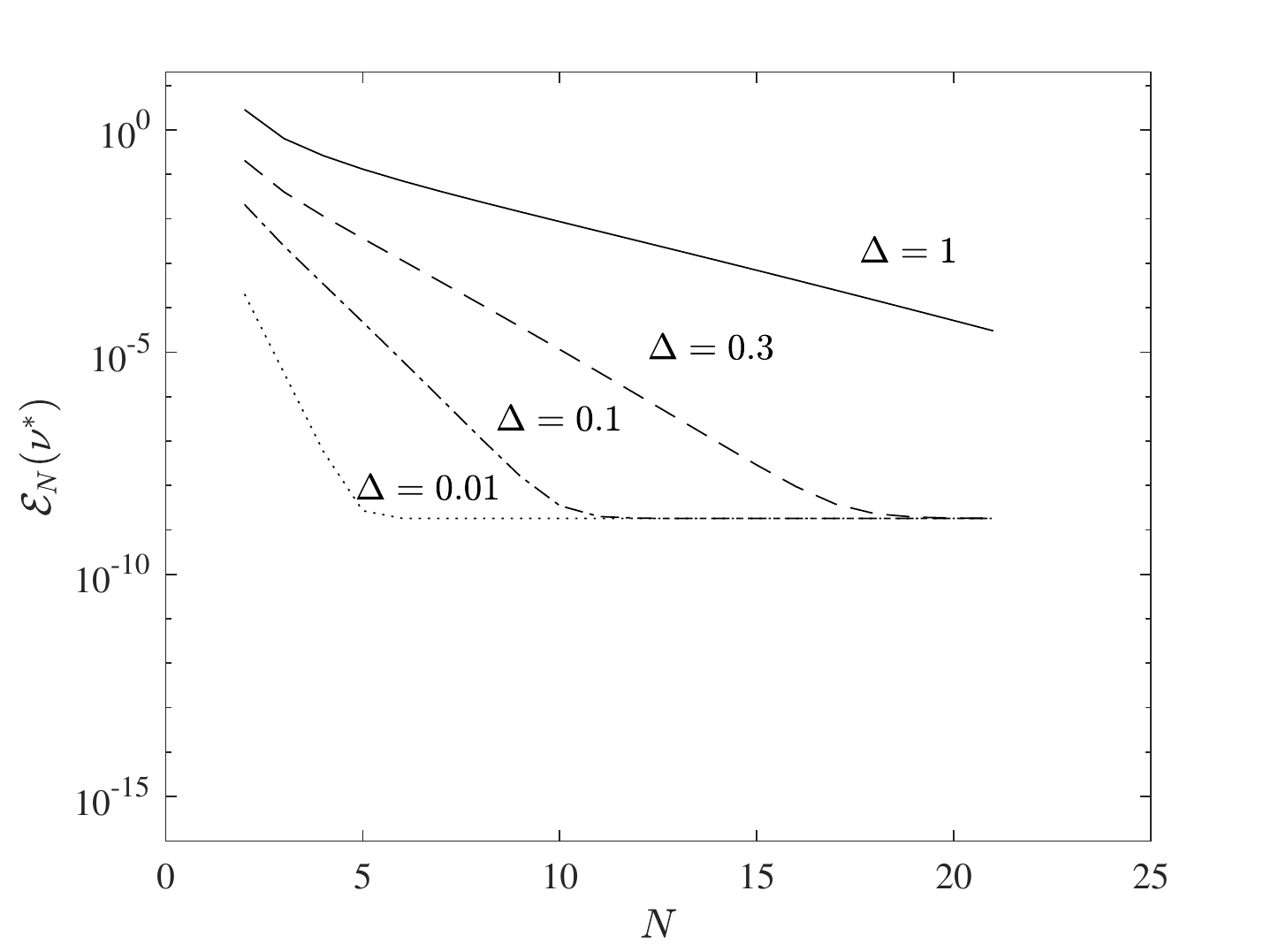}}
 \subfigure[]{\includegraphics[width=0.48\textwidth]{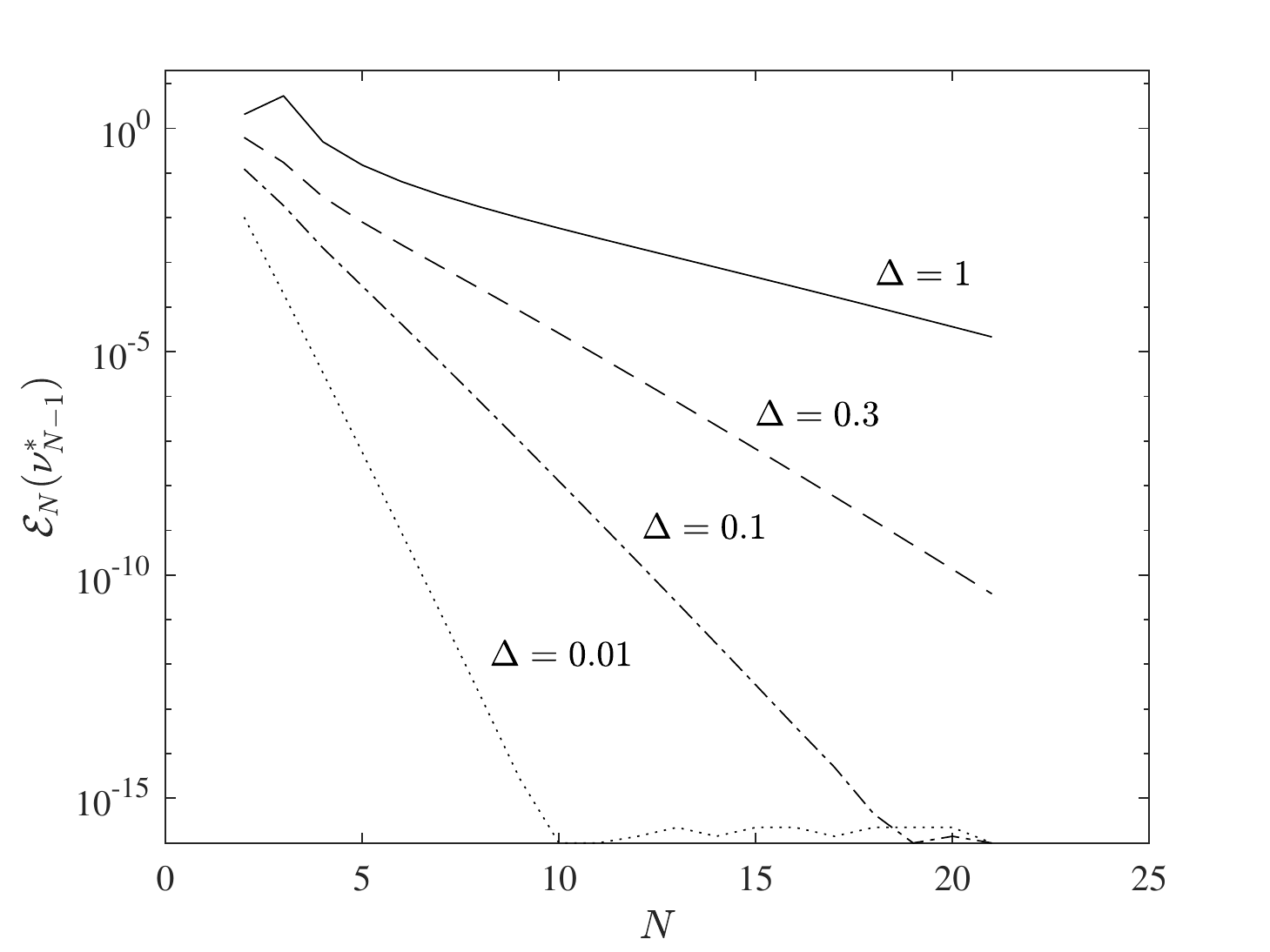}}
  \caption{Convergence of the true error $\mathcal{E}_N(\nu^*)$ (a) and the error estimate $\mathcal{E}_N(\nu^*_{N-1})$ (b) as function of $N$ and the relative distance $\Delta$.}
   \label{fig:ErrEp}
  \end{center}
\end{figure}

The point $\nu_0$ is now chosen in the vicinity of the EP associated with the coalescence of the two least attenuated modes $\lambda_1$ and $\lambda_{2}$. In order to evaluate the robustness and the accuracy of the method different relative distances $\Delta$ are tested.
The relative error defined as 
\begin{equation}\label{eq:Errn}
\mathcal{E}_N(\nu^*)=\frac{| \nu^*_{N}-\nu^*|}{| \nu^* |}
\end{equation}
 is plotted in Fig. \ref{fig:ErrEp}a) as a function of the polynomial order $N$ and for 4 initial distances $\Delta=0.01,\,0.1,\,0.3,\,1$. Here, it is understood that $\nu^* = \nu^*_{1 \rightarrow 2}$ from Eq. \eqref{eq:ZepAna2} and $\nu_N^*$ corresponds to the genuine root of the polynomial $\mathcal{T}_h$.
It can be observed that even for large values of $\Delta$ the method converges to the true EP $\nu^*$.  When the order of the polynomial is increased, the error decreases which shows the stability of the method even in the presence of high-order derivatives. Results are presented up to $N=20$ but some calculation showed that errors of magnitude $10^{-8}$ can be reached for the case $\Delta=1$ when using $N=35$. 
For small values of $\Delta$, a plateau around $10^{-9}$ is quickly reached. This stems mainly from the numerical error due to FEM 
discretization and increasing the number of elements up to 1000 show that the relative error can be as small as $10^{-13}$.
In a more general case, the EP is not known and there is a need for an \emph{a priori} error estimator. 
Here, we propose to use the $N-1$ order  evaluation for the reference value and we consider
$\mathcal{E}_N(\nu^*_{N-1})$ where $\mathcal{E}_N$ given in Eq.~\eqref{eq:Errn}.
It can be observed in Fig.~\ref{fig:ErrEp}b) that the error decreases with the polynomial order $N$
with an exponential slope very similar to the true error estimate $\mathcal{E}_N(\nu^*)$.
However, the error level can reach much smaller values, i.e. near machine precision using standard double precision arithmetic,
and this is due to the fact that, in this case, the solution converges towards the EP of the discrete problem.

\begin{figure}
  \begin{center}
\includegraphics[width=0.48\textwidth]{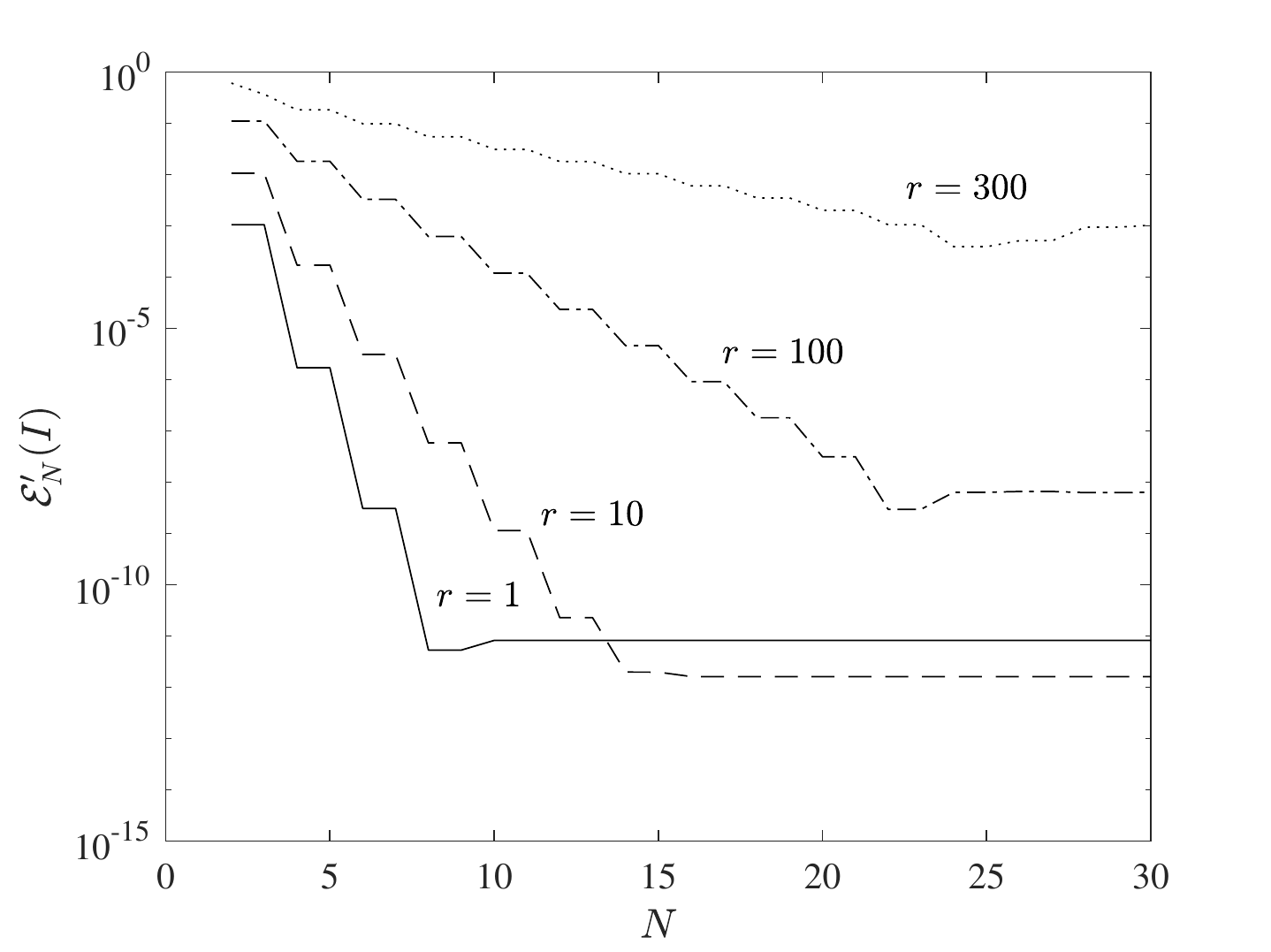}
   \caption{Convergence of the error estimate $\mathcal{E}_N'(I)$ from Eq.~\eqref{errI} with respect to $N$  and for different radius $r$. The relative error on the EP location is around $2\,10^{-9}$.
}
   \label{fig:ErrPuisCoef}
  \end{center}
\end{figure}

In order to investigate the numerical error on the Puiseux series coefficients, let us introduce the contour integral
\begin{equation}\label{eq:Cauchy}
I = \int_\mathcal{C} \lambda(\nu) \ud \nu = \int_{-\pi}^\pi \lambda(r, \varphi)\, \i r  \e^{\i \varphi} \ud \varphi,
\end{equation}
where $\mathcal{C}$ is a circular arc of radius $r$, in the $\nu$-complex plane and centered on $\nu^*$, which avoids the branch cut of the square root principal value.
If $\lambda$ is a regular function within the disk then the integral vanishes by virtue of Cauchy's theorem. 
However, if an EP associated with the multiplicity of $\lambda$ is enclosed,  the contour integral can be obtained from a weighted sum of the Puiseux series coefficients associated with the fractional powers and we have for $\lambda_+$,
\begin{equation}\label{eq:CauchyPuiseux}
I =  \i \sum_{k=1,3,5,\ldots}  (-1)^{\lfloor\tfrac{k}{2}\rfloor} r^{1+\tfrac{k}{2}}  \frac{4}{2+k}  a_k.
\end{equation}
In practice, we shall call $I_N$ the truncated version of Eq. \eqref{eq:CauchyPuiseux} up to order $N$.
The computation of a reference solution of \eqref{eq:Cauchy} is first performed thanks to 
an adaptive integration scheme based on a direct computation of the eigenvalue problem  which can guarantee an absolute precision of around $10^{-12}$ of the value of the integral. During the integration, we ensure to follow the same eigenvalue branch.
The comparison between the direct computation $I$ and the one obtained from the truncated series $I_N$ is shown in Fig.~\ref{fig:ErrPuisCoef}(a), where the error
\begin{equation} \label{errI}
\mathcal{E}_N'(I)=\frac{|I_N-I |}{|I |}
\end{equation}
 is plotted as a function of the number of terms kept in the series for several values of the disk radius $r$. 
 In this example, all Puiseux coefficients have been originally calculated with $\Delta=0.1$ and $N=15$ and the error on the EP location is around $2\,10^{-9}$.
 Curves show `staircase' shapes  because integer powers do not contribute to the contour integral.
When the number of terms  $N$ in the series increases, the error decreases until it reaches a limit which depends on 
the radius and for large values of the latter, the error level is strongly related to the accuracy of high order coefficients of the Puiseux series.

\subsection{Resonant inclusion embedded in a porous lined duct}\label{sec:Bloch3D}

\subsubsection{Finite element model}

\begin{figure}
\centering
\subfigure[Geometry]{\includegraphics[height=8cm]{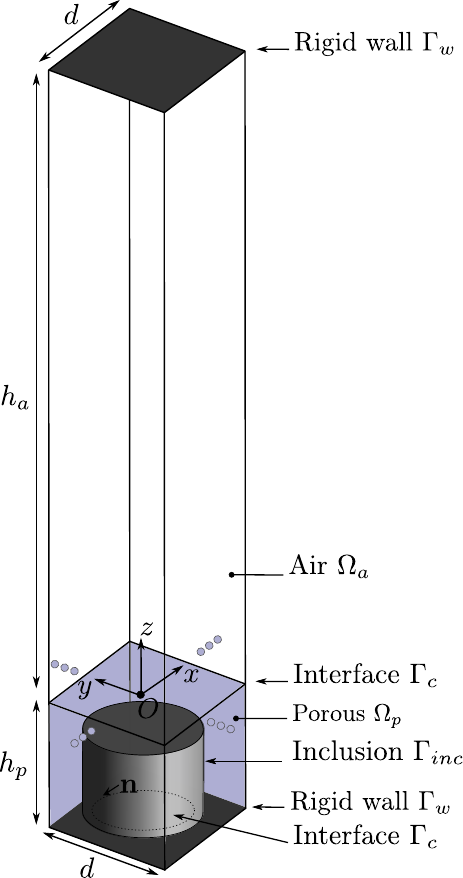}}\hspace{1cm}
\subfigure[mesh \#1 24k dof]{\includegraphics[height=8cm]{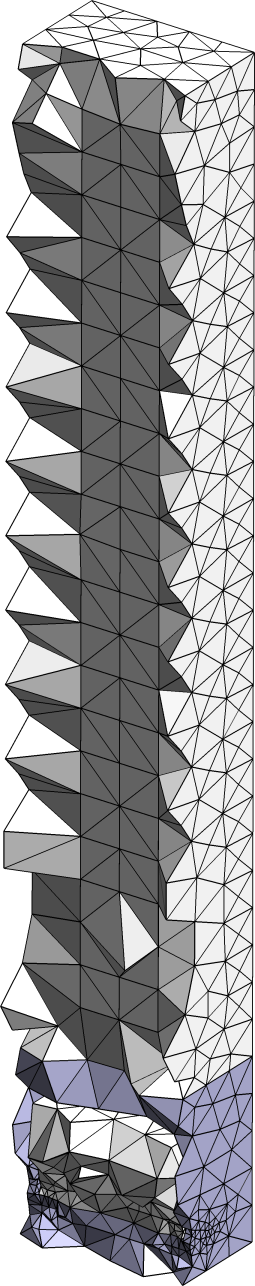}}\hspace{0.5cm}
\subfigure[mesh \#2 176k dof]{\includegraphics[height=8cm]{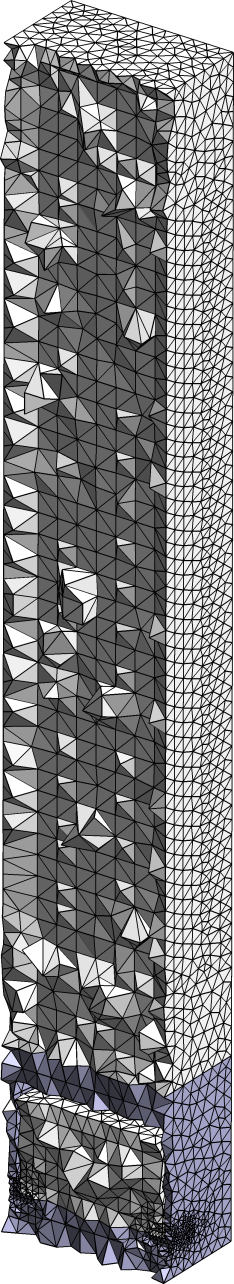}}\hspace{0.5cm}
\subfigure[mesh \#3 300k dof]{\includegraphics[height=8cm]{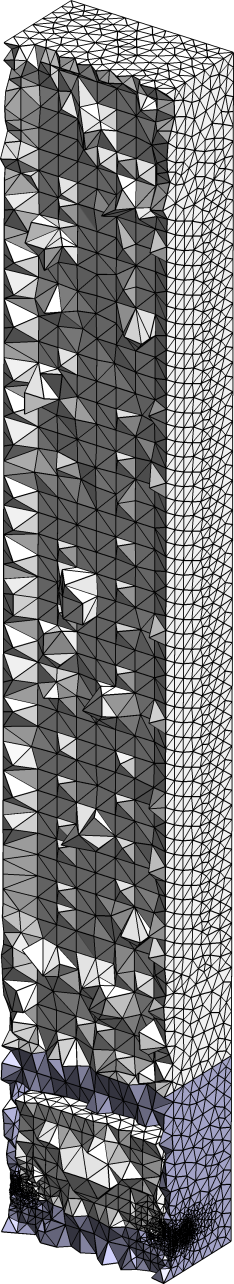}}\hspace{0.5cm}
\subfigure[mesh  \#4 1 071k dof]{\includegraphics[height=8cm]{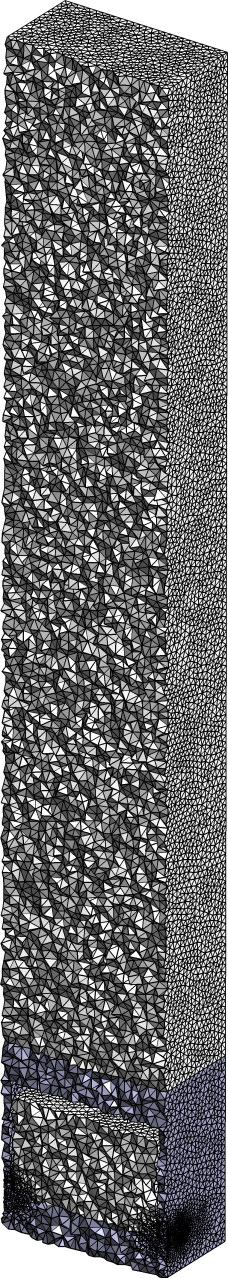}}
\caption{Geometry of the problem adapted from \cite{Bloch3D:2017} colored by physical domain air (white), porous (\textcolor{bluefoam}{blue}) and the meshes used in the computation. Here, $h_a=135$ \si{mm}, $h_p=25$ \si{mm}, $d=24$ \si{mm}.}

\label{fig:schema}
\end{figure}

We consider the propagation of a time-harmonic acoustic wave  ($\e^{-\i \omega t}$ convention is adopted here) in a 3D infinite guiding structure which is periodic both in the $x$ and $y$-direction. A typical unit cell is illustrated in Fig.~\ref{fig:schema} where the air domain $\Omega_a$ of height $h_a$ is  lined with a porous material $\Omega_p$ of thickness $h_p$ with an embedded rigid cylindrical inclusion of circular shape filled with air. This configuration has already been studied in \cite{Bloch3D:2017}.
We call $\Gamma_c$ the two interfaces between the air and the porous material whereas $\Gamma_{inc}$ and $\Gamma_{w}$ stands respectively for the surface of the inclusion  and the rigid walls of the waveguide.
The dimensions of the  rectangular shaped cell are: the total height $H=h_a+h_p$ and the width $d$ wich is identical in both $x$ and $y$-direction. 
Thus the guiding structure consists of a regular $\mathbf{d}$-periodic grid with $\mathbf{d}=[d,\,d,\, 0]^ \mathrm{t}$.
The skeleton of the porous material is supposed to be infinitely rigid, thus the use of the Johnson-Champoux-Allard (JCA) equivalent fluid model~\cite{Allard:1993} is appropriate and we call $K_p(\omega)$ the equivalent bulk modulus  and $\rho_p(\omega)$  the density (see Appendix~A in \cite{Bloch3D:2017} for details). The sound speed in the porous material is given by $c_p(\omega) = \sqrt{K_p(\omega)/\rho_p(\omega)}$. Finally, for the air domain we took the standard values:  $\rho_a=1.213$ kgm$^{-3}$, $K_a= 141\,855$ Pa and the sound speed is given by $c_a=\sqrt{K_a/\rho_a}$. 
In this parametric study, we assume that the complex density $\rho_p$ can vary independently of the other parameters, i.e. $K_p$, $\rho_a$, $K_a$ and $\omega$
which are supposed constant. 

In each domain $\Omega_{\alpha}$ ($\alpha=a,p$), the pressure field is governed by the Helmholtz equation
\begin{equation}\label{eq:Helmholtz2}
\Delta p_{\alpha}  + k_{\alpha}^2 p_{\alpha} = 0,
\end{equation}
where $k_{\alpha}=\omega/c_{\alpha}$ stands for the wavenumber in both domains.
Air and porous media are coupled together at the interface $\Gamma_c$  where continuity of pressure, i.e. $p_a=p_p$, and of the normal velocity, i.e. 
\begin{equation}\label{eq:veln}
\frac{1}{\i \omega \rho_a}\nabla p_a \cdot \n = \frac{1}{\i \omega \rho_p}\nabla p_p \cdot \n,
\end{equation}
must be satisfied. On rigid walls ${\Gamma}_w$ and ${\Gamma}_{inc}$, the normal velocity vanishes and $\nabla p_\alpha \cdot \n  = 0$.
Thanks to Bloch theorem, general solutions of are expressed in terms of Bloch waves of the following form
\begin{equation}\label{eq:Bloch}
p_\alpha(\xv) = \phi_\alpha(\xv) \e^{\i \kv_B \cdot \xv},
\end{equation}
which means that the pressure field is written as the product of a $\mathbf{d}$-periodic field $\phi_\alpha(\xv)$  modulated by a plane wave traveling in the direction given by the Bloch wavevector $\kv_B$. The real part of $\kv_B$ measures the change in phase across the cell and its imaginary part the attenuation. 
It is adequate to put $\kv_B= k_B \kappav$ where the unit vector $\kappav=\kv_B/k_B$ stands for the direction of propagation and $k_B$ is the wavenumber.  In the example treated here, we are looking to solutions traveling along the $x$-direction, thus $\kappav =[1,\,0,\, 0]^ \mathrm{t}$.
The problem can now be rewritten  using the periodic field $\phi$, i.e.
\begin{equation}\label{eq:HelmholtzP}
 \Delta \phi_\alpha + 2\i k_B \kappav \cdot \nabla \phi_\alpha  + (k_\alpha^2- k_B^2)\phi_\alpha  = 0.
\end{equation}
The associated weak formulation is obtained by multiplying Eq.~\eqref{eq:HelmholtzP}
by a periodic test function $\psi_\alpha$ ($\alpha=a,\,p$) and, after integration by parts over a unit cell,  we obtain 
\begin{multline}
\sum_{\alpha=a,p} \frac{1}{\rho_\alpha} \times  \biggl\lbrace   - \int_{\Omega_\alpha} \nabla \psi_\alpha \cdot \nabla \phi_\alpha  \ud \Omega + k_\alpha^2 \int_{\Omega_\alpha}  \psi_\alpha \, \phi_\alpha\, \ud \Omega   
 + \i k_B \int_{\Omega_\alpha} \Big( - \nabla \psi_\alpha  \cdot (\kappav \phi_\alpha )  +  (\kappav  \psi_\alpha )\cdot  \nabla \phi_\alpha \Big) \ud \Omega \\
 - k_B^2 \int_{\Omega_\alpha}  \phi_\alpha\,\psi_\alpha   \ud \Omega 
+ \int_{\partial \Omega_\alpha} \psi_\alpha \left(\frac{\partial \phi_\alpha}{\partial n} + \i k_B (\kappav \cdot \n) \phi_\alpha  \right)  \ud \Gamma \biggr\rbrace =0.
\end{multline}
In this work, the variational formulation is discretized using standard quadratic tetrahedral Lagrangian finite elements.
In order to preserve the block matrices structure associated with each domain thus avoiding  the need for renumbering, 
continuity conditions across the air-porous interface and periodicity are taken into account using Lagrange multiplier techniques. 
The density of the fluid and the material appearing explicitely in the equation above permits to ease the treatment 
of the continuity of the normal velocity at the interface.
This finally leads to a quadratic eigenvalue problem which can be written in the general form
\begin{equation}\label{eq:polynomKB}
\Lv \left( \lambda (\nu),\nu \right) \xv (\nu) = \left[\Kv_0(\nu) + \Kv_1(\nu) \lambda + \Kv_2(\nu) \lambda^2 \right] \xv (\nu) =0.
\end{equation}
In order to be consistent with our notations, we put $\lambda=k_B$ and the parameter of our problem is chosen to be the density of the porous material, $\nu=\rho_p$ which, in the present context must be regarded as an independent variable and this means that the quantity $k_p^2/\rho_p$ does not depend on the parameter $\nu$.
Note that the sound absorbing properties of the porous material are related to the fact its physical parameters are all complex-valued and none of the matrices $\Kv_i$ ($i=0,1,2$) are  Hermitian but either complex symmetric or complex skew symmetric. Here, the explicit dependence 
of these matrices as function of $\nu$ is deliberately not explicity written 
in order to be very general in the presentation of the method.
Applying general Liebnitz' rules for derivative of product to the original eigenvalue problem \eqref{eq:PVP} yields
\begin{multline}
\frac{\ud^n}{\ud \nu^n}  \left[\Lv \left( \lambda (\nu),\nu \right) \xv (\nu) \right]  = \\ \sum_{k=0}^n  \binom{n}{k} \Kv_0^{(k)} \xv^{(n-k)}
+
\sum_{k_1+k_2+k_3=n} {n \choose k_1, k_2, k_3} \left[\Kv_1^{(k_1)} \xv^{(k_2)} \lambda^{(k_3)} + \Kv_2^{(k_1)} \xv^{(k_2)} \left(\lambda^2\right)^{(k_3)} \right] =0 ,
\end{multline}
where the sum extends over all $3$-tuples ($k_1,k_2,k_3$) of non-negative integers  and $ {n \choose k_1, k_2, k_3}
 = \frac{n!}{k_1!\, k_2! k_3!}$ are the multinomial coefficients. 
The successive derivative of the squared eigenvalue is obtained using Liebnitz' rule 
\begin{equation}
\left(\lambda^2\right)^{(k_3)} = \sum_{k=0}^{\lfloor \tfrac{k_3}{2}\rfloor} \binom{k_3}{k} \, \left(2-\delta_{\tfrac{k_3}{2}k}\right)   \lambda^{(k_3-k)}\lambda^{(k)}.
\end{equation}

\noindent The right hand side in Eq.~\eqref{eq:Bordered} is computed explicitely using
\begin{align}\label{eq:Bloch3DRec}
\Fv_n &= \frac{\ud^n}{\ud \nu^n}  \left[\Lv \left( \lambda (\nu),\nu \right) \xv (\nu) \right] - \left(\Lv \xv^{(n)} + \partial_\lambda \Lv \xv \lambda^{(n)}\right) \nonumber\\
 &= \sum_{k=1}^n  \binom{n}{k} \Kv_0^{(k)} \xv^{(n-k)}
+
\sum_{\substack{k_1+k_2+k_3=n\\ k_2\neq n , k_3 \neq n}} {n \choose k_1, k_2, k_3}\Kv_1^{(k_1)} \xv^{(k_2)} \lambda^{(k_3)}  
\\ 
& \quad + \sum_{\substack{k_1+k_2+k_3=n\\ k_2\neq n}} {n \choose k_1, k_2, k_3}\Kv_2^{(k_1)} \xv^{(k_2)} \sum_{k=\delta_{k_3n}}^{\lfloor \tfrac{k_3}{2}\rfloor} \binom{k_3}{k} \, \left(2-\delta_{\tfrac{k_3}{2}k}\right)   \lambda^{(k_3-k)}\lambda^{(k)}. \nonumber
\end{align}
These expressions, though cumbersome, can be formally derived  automatically with symbolic methods.
The fact that block matrices structure has been been preserved allows to compute derivatives of $\Kv_i$ analytically and
this also facilitates the computation of Eq.~\eqref{eq:Bloch3DRec}. Note that about half of entries of matrices depends on the parameter, and, in contrast with the example treated in the previous section where only a single entry was involved through $\boldsymbol\Gamma$.

\subsubsection{Results} \label{sec:BlochResults}
\begin{figure}
  \begin{center}
\subfigure[EP corresponding to the coalescence of  $(\lambda_1,\lambda_2)$, $\nu_0=1.+0.1\i$. Here $\nu^* \in \lbrace 0.937+0.441\i,\, 0.889-0.599\i \rbrace$. ]{\includegraphics[width=0.48\textwidth]{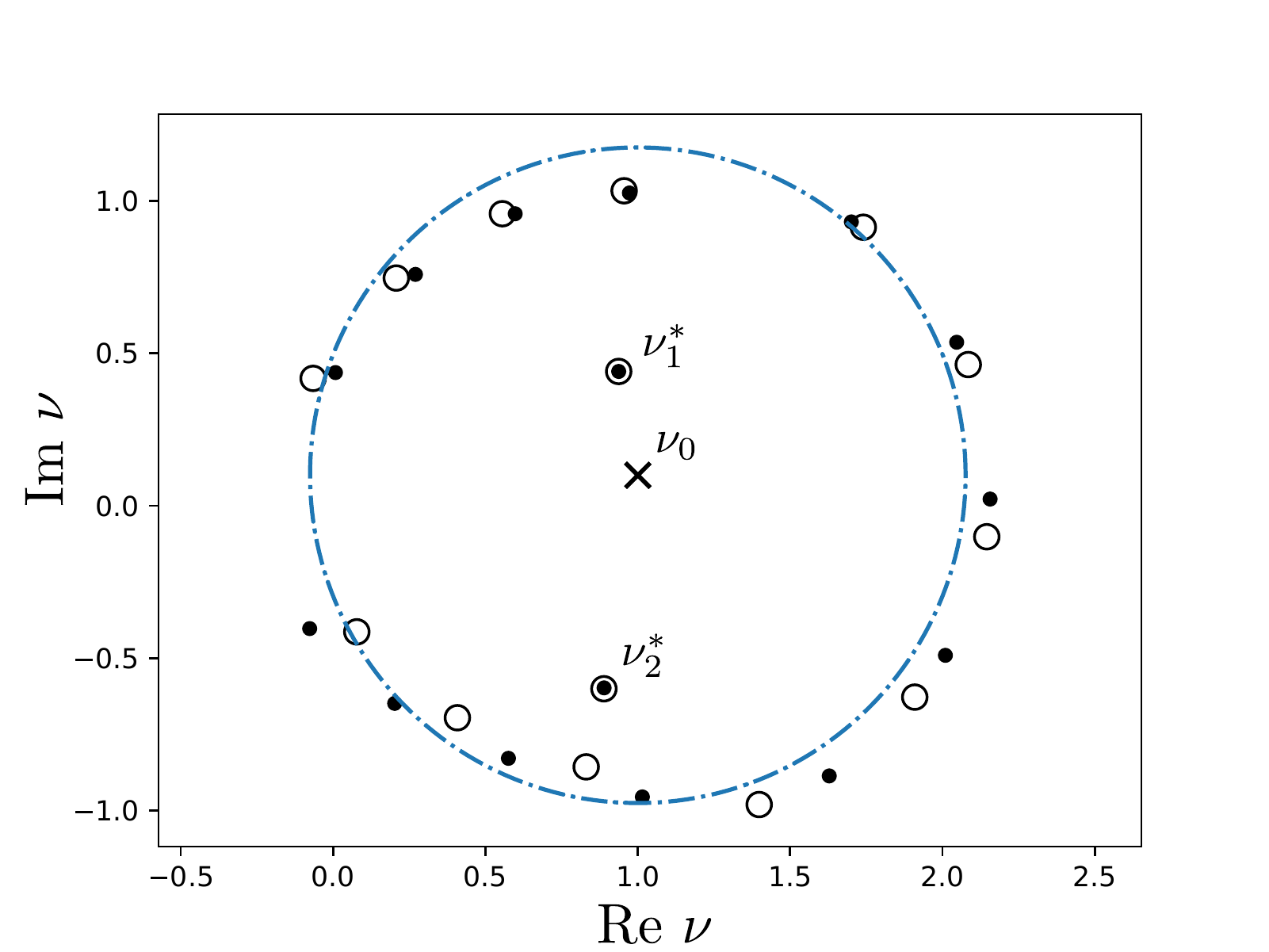}}
\subfigure[EP corresponding to the coalescence of  $(\lambda_2,\lambda_3)$, $\nu_0=2.+0.1\i$. Here $\nu^* \in \lbrace 1.201+0.891\i,\, 1.199-0.984\i\rbrace$ ]{\includegraphics[width=0.48\textwidth]{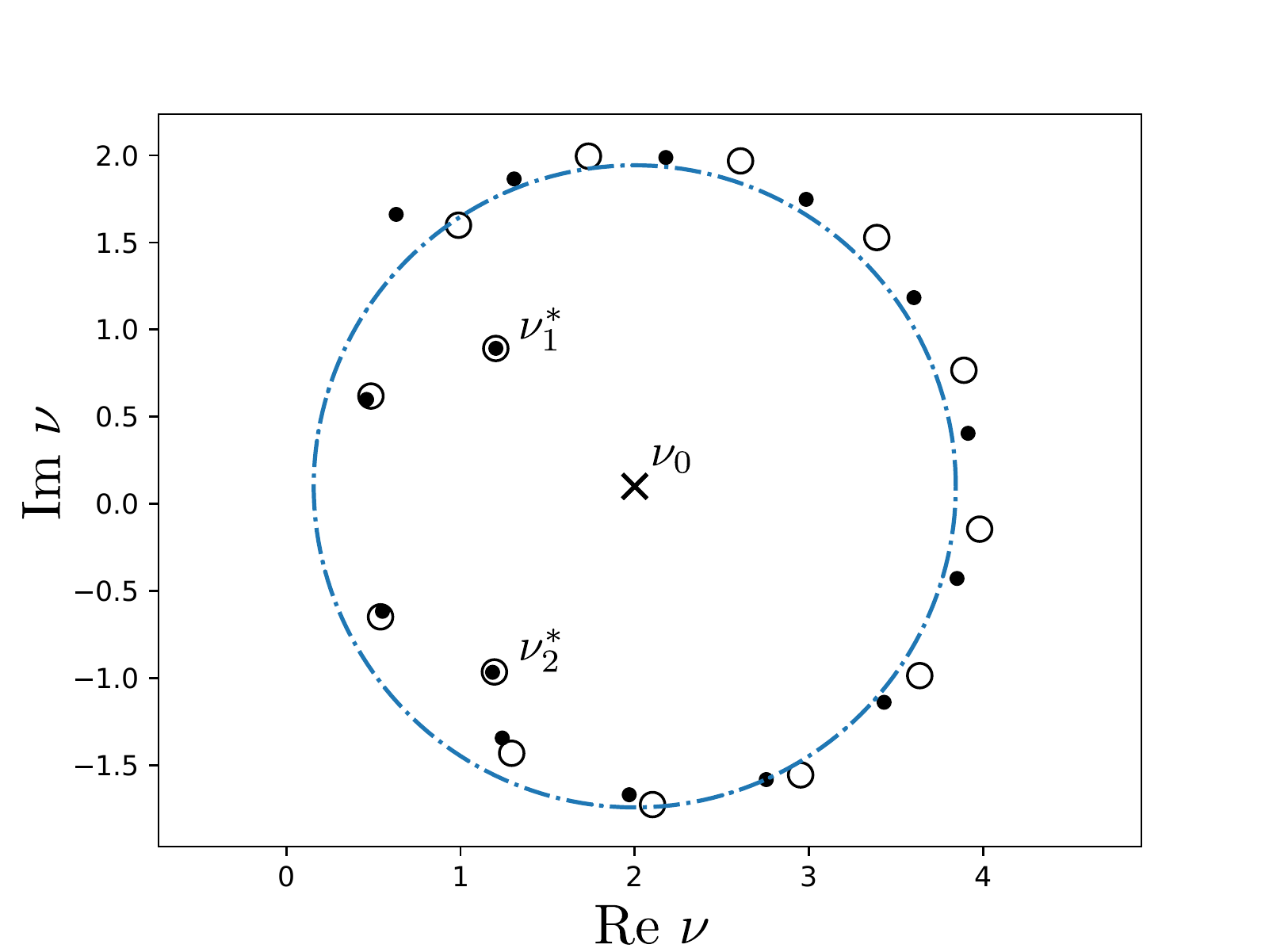}}
%
   \caption{EP localization for two selected eigenvalues, computed with mesh \# 1 and $N=14$. Here, the circle (\textcolor{blue}{\LegendDashdot})
   	 is centered on $\nu_0$ with radius given by Eq.~\eqref{eq:rootradius}, symbols $\bullet$ stands for the roots of $\mathcal{T}_h^N$ and $\circ$ the roots of $\mathcal{T}_h^{N-1}$. }
   \label{fig:EPpairs}
  \end{center}
\end{figure}

The method is applied for the configuration given in Fig.~\ref{fig:schema} discretized with mesh \#1. 
The porous material is a metallic foam with the following JCA parameters : porosity $\phi=0.99$, tortuosity $\alpha_\infty=1.17$, viscous characteristic length $\Lambda=1\times 10^{-4}$ \si{m}, thermal characteristic length $\Lambda'=2.4\times 10^{-4}$ \si{m}, and air flow resistivity $\sigma=6900$ \si{N.m^{-4}.s}.
We are interested in locating the EP  associated with the coalescence of the two pairs of eigenvalues among the three least attenuated mode, i.e. the pair ($\lambda_1$, $\lambda_2$) and ($\lambda_2$, $\lambda_3$). The roots of the truncated Taylor series $\mathcal{T}_h$ are given in both cases in Fig.~\ref{fig:EPpairs}a) and in Fig.~\ref{fig:EPpairs}b) respectively. 
It can be observed that the roots, as in the previous section, tend to be aligned on a circle and genuine zeros corresponding to the EPs are located in the circle. A simple way to discriminate between genuine and spurious roots is to compute the roots of $\mathcal{T}_h$ for two successive orders.
This is illustrated in Fig.~\ref{fig:EPpairs} where roots are computed by taking $N= 14$ and $N= 15$. In contrast with genuine roots, the location of the spurious roots depend strongly on the approximation order, as expected since spurious roots generally tend to be evenly distributed on the circle.
In this specific case shown here, two EPs are found at $\nu^* \in \lbrace 0.937+0.441\i,\, 0.889-0.599\i \rbrace$ for the pair ($\lambda_1,\lambda_2$) and at $\nu^* \in \lbrace 1.201+0.891\i,\, 1.199-0.984\i\rbrace$ for the pair ($\lambda_2,\lambda_3$). 
The algorithm is able to identify 2 different EPs which signifies that the same pair of eigenvalues merge at two different locations. 
Note that these two values are almost complex conjugate and this stems from the nature of the original problem. Though this is not the place for a rigorous analysis of the matter, some simple arguments can be brought forward. First, any operator satisfying 
$\overline{\Lv ( \lambda,\nu)}= \Lv ( \bar{\lambda},\bar{\nu})$ admits EPs which are necessarily complex conjugate and this is reflected, for instance,
by the numerical example shown in the previous section (it suffices for this to consider the 'admittance' $\nu = \i/Z$, instead of $\nu=Z$, in Eq. \eqref{eq:PVPz}). Second, the fact that 
this associated pair of EPs stem from the coalescence of the same pair of eigenvalues can be understood in the ideal scenario where 
Taylor coefficients of function $h$ are all real-valued. Again, this happens in  Eq. \eqref{eq:PVPz} whenever the initial value $\nu_0$ for the computation is chosen on the real axis.

\begin{figure}
  \begin{center}
\subfigure[Real part of $(\lambda_1,\,\lambda_2)$]{\includegraphics[width=0.32\textwidth]{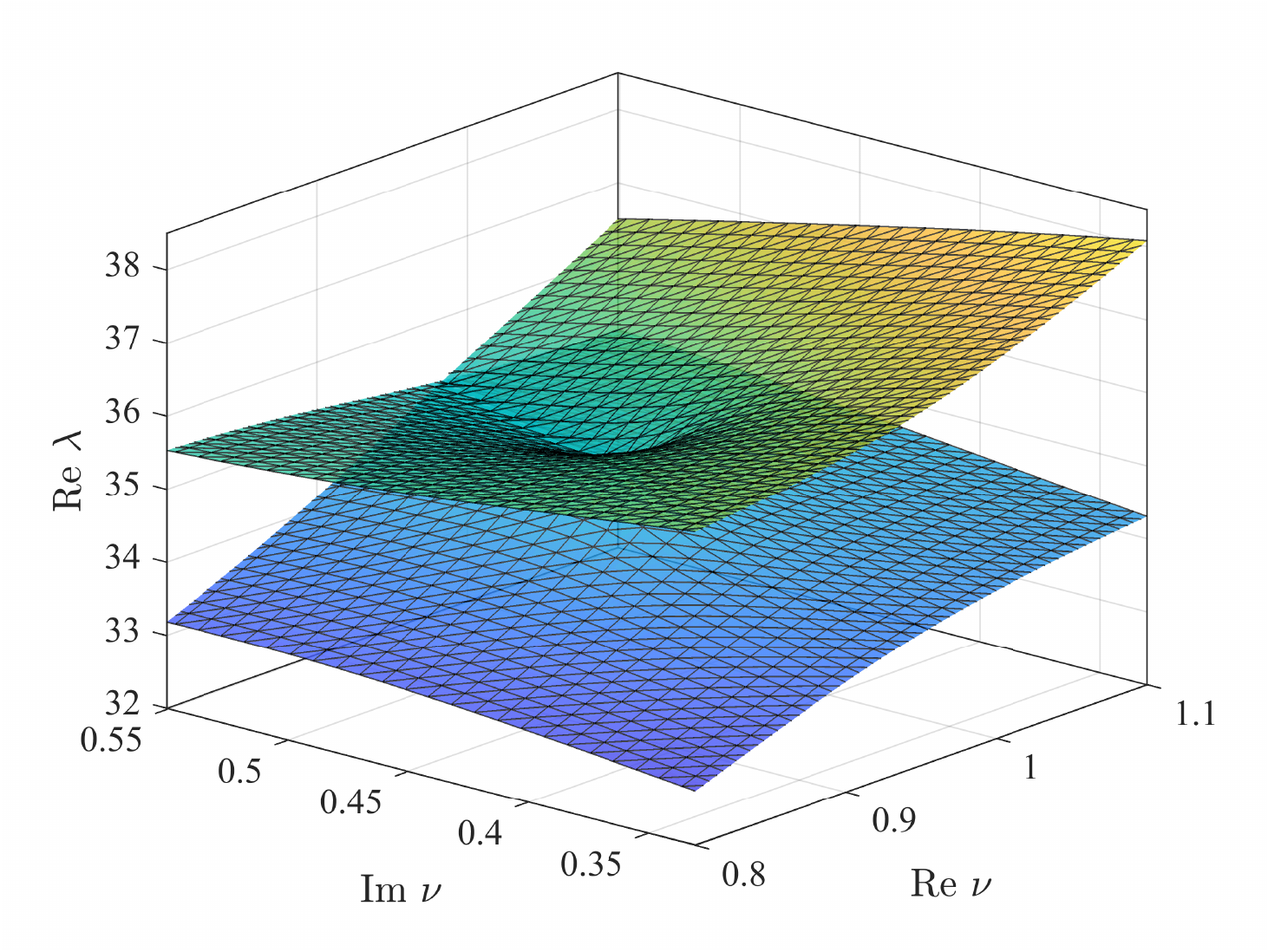}}
\subfigure[Imaginary part of $(\lambda_1,\,\lambda_2)$]{\includegraphics[width=0.32\textwidth]{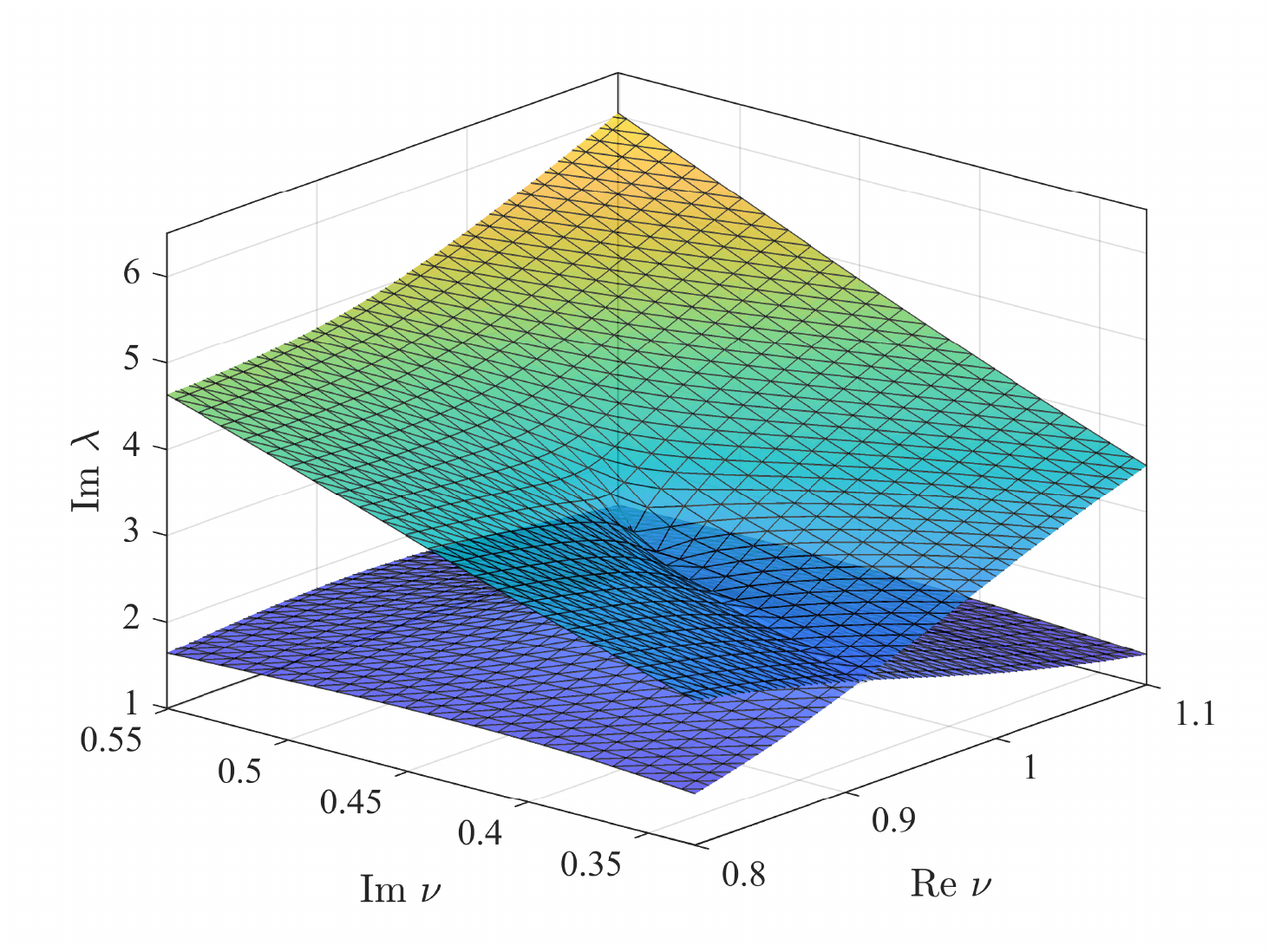}}
\subfigure[Error level $\log_{10}\mathcal{E}_\mathcal{P}$]{\includegraphics[width=0.32\textwidth]{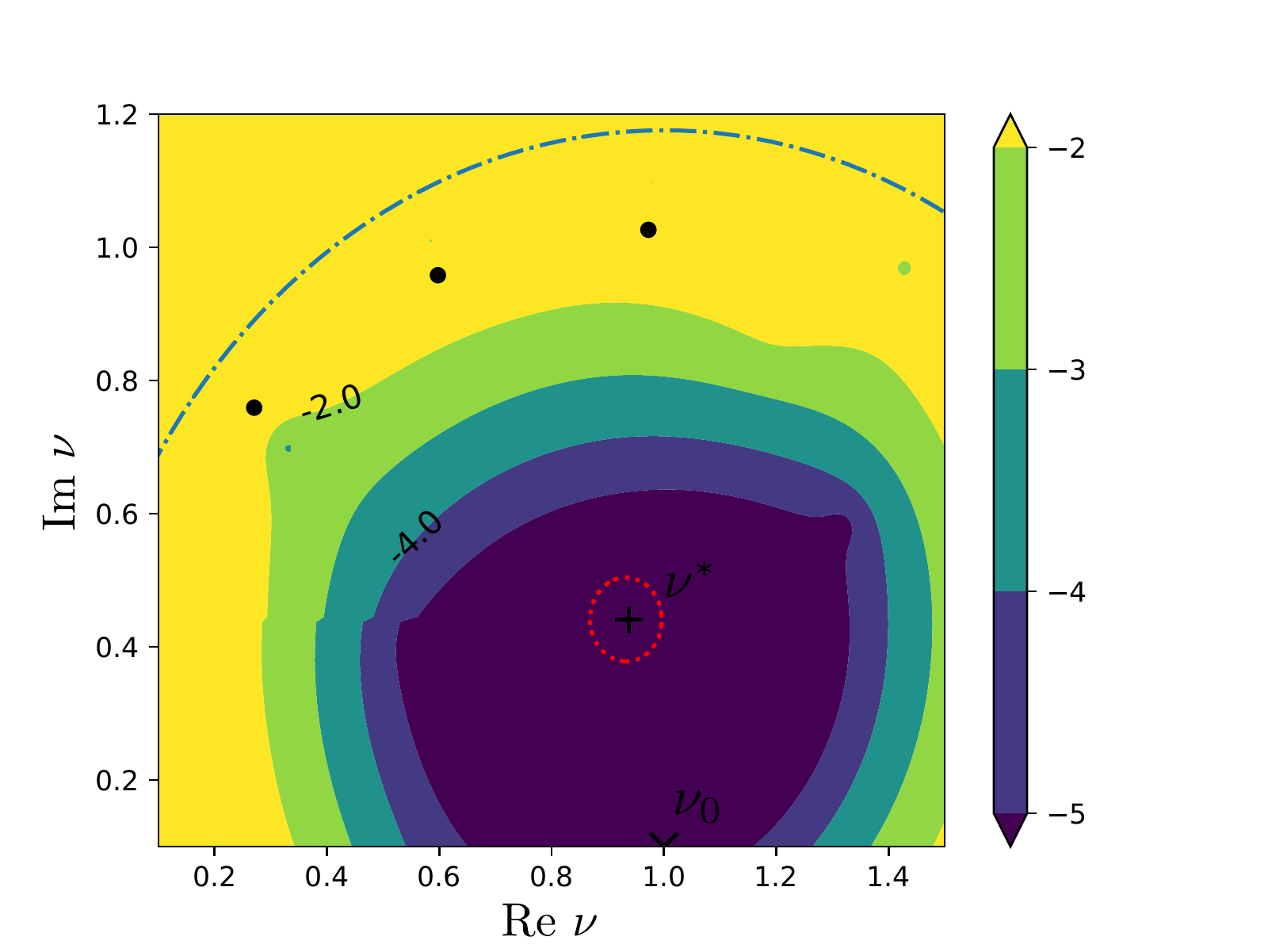}}
   \caption{Direct computation and truncated Puiseux series with $N=14$ showing the real and imaginary parts of two coalescing eigenvalues in the vicinity of an exceptional point $\nu^*=0.937+0.441\i$. 
   	Differences are shown in logarithmic scale  and the red dotted line indicates the level of error, here $1 \%$, if the Puiseux series is truncated after the square root term.}
   \label{fig:RiemannPuiseux}
  \end{center}
\end{figure}

In Fig.~\ref{fig:RiemannPuiseux} are plotted the real and imaginary parts of two merging eigenvalues in the neighborhood of the EP $\nu^*=0.937+0.441\i$, obtained either by operating a direct eigenvalue computation or by using the truncated Puiseux series. The computed series is given explicitely here
\begin{multline}
\lambda_\pm \approx 
  (35.114 + 2.8269\i)
\pm (2.8088 + 3.8317\i) (\nu-\nu^*)  ^\frac{1}{2} 
+ (5.0539 + 1.4549\i) (\nu-\nu^*) \\
\pm (2.2725 + 1.0034\i) (\nu-\nu^*)^\frac{3}{2} 
+ (2.3618 + 2.2925\i)(\nu-\nu^*)^2 
\pm (4.0180 + 0.4889\i) (\nu-\nu^*)^\frac{5}{2} 
+ \dots
\end{multline}
An excellent agreement is observed as there are no discernible differences between the reference solution and the one recovered by the Puiseux series.
In Fig.~\ref{fig:RiemannPuiseux}c) the error level (in logarithmic scale) is shown. It can be seen that  the truncated Puiseux series is accurate 
over a relatively large region limited by the radius of convergence of the Taylor expansion of $h$ with respect to $\nu_0$. Beyond this region, indicated by a dash-dotted circle, there exists another EP involving one eigenvalue among the selected pair and this latter could be recovered using another Puiseux series.
The fact that EPs are associated with optimal attenuation can be understood 
from a graphical point a view as the two eigenvalues, here $\lambda_1$ and $\lambda_2$ which correspond to the two least attenuated modes,  coalesce in opposite directions where the imaginary part is expected to reach a maximum, at least locally and this is clearly illustrated in Fig.~\ref{fig:RiemannPuiseux}b). At the EP, the spatial representation of the acoustic pressure mode in the unit cell exhibits
strong localisation of the acoustic energy in the porous material as shown and discussed in the recent paper~\cite{Bloch3D:2017}. Note that a similar study could be conducted for the other EP associated with a density with negative imaginary part, here $\nu^*=0.889-0.599\i$, and in this situation, the material has  to be regarded as active though the physical existence of such material remains an open question.

\subsection{Computational efficiency}

The computational time of the proposed algorithm depends on different steps, which are reported in Algorithm~\ref{al:global}.
All steps are automated and the overall accuracy is mainly controlled by the quality of the original mesh (i.e. the matrix size $\mathcal{N}$) and $N$ the number of terms in the truncated Taylor series. In practice, choosing $N$ between 8 and 14 has been found to be sufficient. The last parameter to set is the tolerance $\mathcal{E}_N(\nu_{N-1}^*)<$\texttt{tol}. This is used to discriminate true EP from spurious roots.
The type of problems we are solving involve large size sparse matrices, which is somehow standard for 3D geometries. Thus the cost for the computation of the roots and the  Puiseux series coefficients
becomes negligible and most of the computational burden stems from the eigensolver and the LU decomposition of the bordered matrix.
In order to assess the scalability of the proposed method with respect to the number of processors used in the calculation, the coarse mesh \# 1 in Fig.~\ref{fig:schema}b) is progressively refined and the computational work is carried out using parallel computing on multiple processors, here from 1 to 12. The parallelism is driven by PETSc  \cite{Petsc} and concern all matrix operations, linear solve from \cite{Mumps} and the eigenvalue computation  \cite{SLEPc}. In all cases, 10 eigenvalues corresponding to the least attenuated acoustic modes of the guiding structure are considered and the first 10 coefficients of the Puiseux series are recovered.
The time for the main steps of computation are reported in the columns of Tab.~\ref{tab:time}. These are:  the eigenvalue computation, the LU factorization of the bordered matrix from Eq.~\eqref{eq:Bordered}, the forward and backward substitution to solve  Eq.~\eqref{eq:Bordered} with multiple right-hand side vectors $\Fv_n$ ($n=1,\ldots,N$) and the time required to compute the RHS. The latter is constructed using the recursive formula \eqref{eq:Bloch3DRec} and, as expected, the time grows with the number of terms in the Taylor series. In the second last  column is reported the total computational time of the method.
More importantly, this time must be compared to the time spent for the computation of eigenvalues for a specific value of the parameter, here $\nu_0$. In order to illustrate this, the last column shows the time ratios. It is interesting to see that this ratio is relatively stable, say between 2 and 4, whatever the configuration. 
On the basis of these observations, it is clear that the method developed in this work clearly outperforms 
other techniques based on low order Puiseux series like the 3 points methods \cite{Uzdin:2010} or the octahedron method \cite{Cartarius:2016}.
For large size problems, most of the computational burden stems from one single eigenvalue computation and two LU factorization, one for each 
selected eigenvalue, and the ratio is close to 2. Using several processor allow to efficiently reduce the computational time of these steps.

\begin{figure}
  \begin{center}
\includegraphics[width=0.45\textwidth]{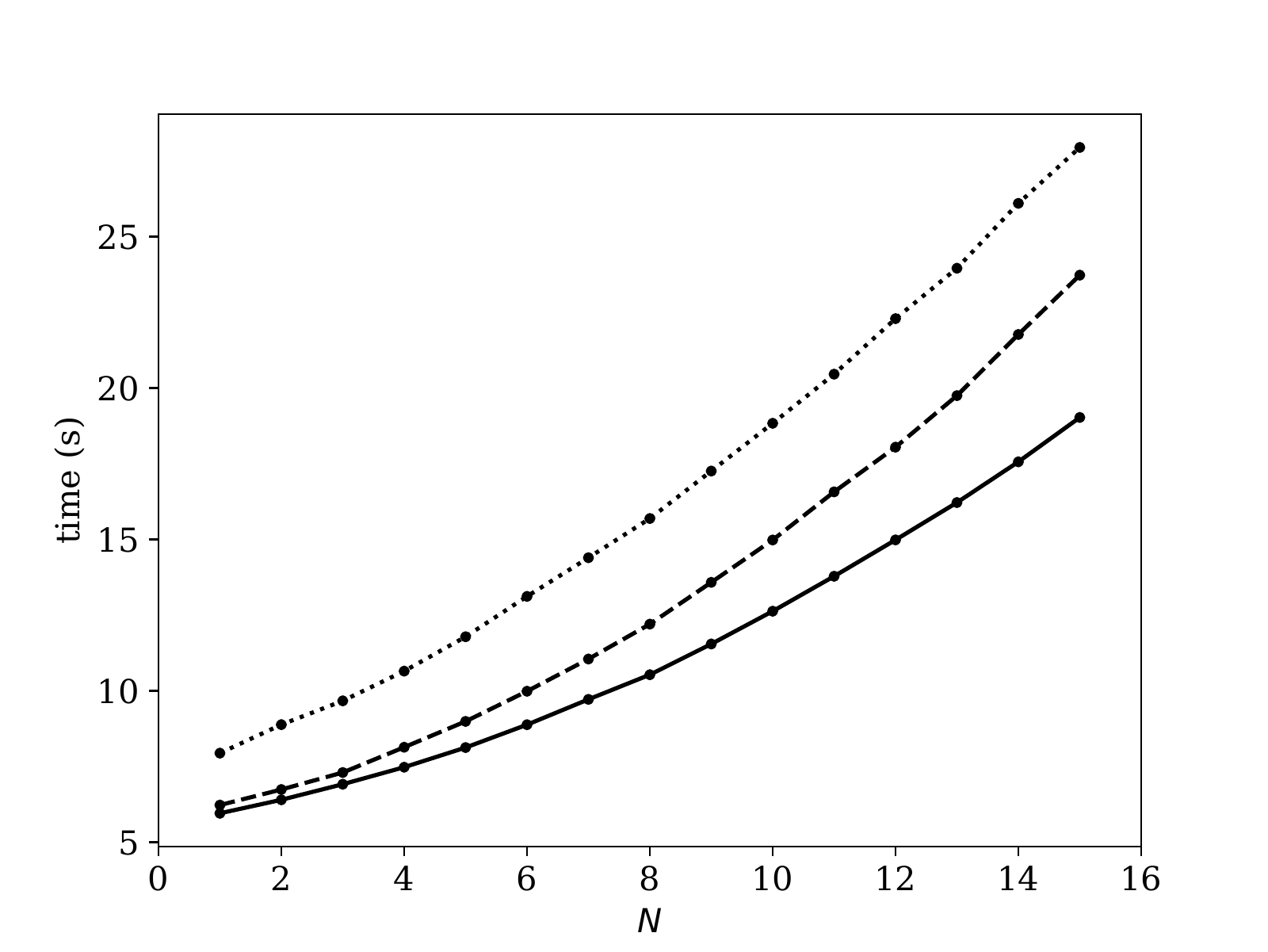}
   \caption{Evolution of the RHS building time according to the requested number of derivative $N$ for the refined  mesh \#4 with 4 (\LegendDotted), 8 (\LegendDashed) and 12 (\LegendLine) CPUs.}
   \label{fig:rhsTime}
  \end{center}
\end{figure}

Finally, though negligible for large size matrices, it is worth investigating the computational time spent for the construction of the right-hand side vector $\Fv_n$.
It can be observed in Fig.~\ref{fig:rhsTime} that this time tends to grows quadratically with the number of derivatives $N$. This is in line with the fact that the example treated here involves the derivative of the product of 3 functions, and the number of terms which appears after application of Leibniz's rule can be shown to behave asympotically like $N^2$. 
For the problem treated in section \ref{sec:impedance}, 2 functions are involved and  the computational time only grows linearly with $N$.
Here, the parallelism used for the matrix-vector products allows significant time reduction.



\begin{algorithm}[htb]
\algrenewcommand\algorithmicrequire{\textbf{Input:}}
\algrenewcommand\algorithmicensure{\textbf{Output:}}
\algrenewcommand{\algorithmiccomment}[1]{\hfill{\color{black!60}$\triangleright${#1}}}
\begin{algorithmic}[1]
\Require Starting guess $\nu_0$, the discrete operator $\Lv$ and its  derivatives, a tolerance $\mathtt{tol}$ and the number of terms $N$
\Ensure Localisation of the exceptional point $\nu^*$ and Puiseux series coefficients $\mathbf{a}$ 
\State Solve the eigenvalue problem \eqref{eq:PVP} of size $\mathcal{N}\times\mathcal{N}$
\State Select a pair of near merging eigenvalue ($\lambda_+$,$\lambda_-$)
\State LU decomposition of the bordered matrix $(\mathcal{N}+1)\times(\mathcal{N}+1)$ from Eq.~\eqref{eq:Bordered}  for each eigenvalue ($\lambda_+$,$\lambda_-$)
\State Compute iteratively $\lambda_\pm^{(n)}(\nu_0)$ $n=1,\ldots,N$  
\State Compute the coefficients of the truncated taylor series $\mathcal{T}_h$ from Eq. \eqref{eq:dh}
\State Compute the complex roots $\zeta_n$ of $\mathcal{T}_h$ with a companion matrix of size $N\times N$
\State Identify the circle of convergence centered on $\nu_0$ with 
radius given by Eq. \eqref{eq:rootradius}
\If{roots are inside the circle} \Comment{There are some EP candidate}
	\For{each root}
	    \State Compute the error estimator $\mathcal{E}_N(\nu_{N-1}^*)$ from Eq. \eqref{eq:Errn}
	    \If{$\mathcal{E}_N(\nu_{N-1}^*)< \mathtt{tol} $} \Comment{the root is an EP}
        		\State Compute $\mathcal{T}_g$ Taylor expansion from Eq. \eqref{eq:dg}
        		\State Compute Puiseux series coefficients $\mathbf{a}$ with Eqs. \eqref{eq:ae} and \eqref{eq:ao}
			\State Store ($\nu^*$, $\mathbf{a}$)
        	\EndIf 
   \EndFor
   \State \Return ($\nu^*$, $\mathbf{a}$)-list if any
\Else \Comment{No EP found in the $\mathcal{T}_h$ convergence disk}
        \State \Return Choose a new $\nu_0$ and start again
\EndIf 

\end{algorithmic}
\caption{EP Localization and computation of the coefficients of the Puiseux series.}\label{al:global}
\end{algorithm}

\begin{table}
\begin{center}
\begin{tabular}{ccccccc}
\hline
$\mathcal{N}(\#CPU)$ & Eig. & LU & FB sub. & Built RHS & \multicolumn{2}{c}{Total time for 10 Derivatives } \\
\cline{6-7}
- & s & s & s & s & s & /Eig.  \\
\hline
24 696 (4) & 1.21 & 0.73 & 0.02 &   [ 0.11,   0.21 ] & 4.85 & 4.00 \\
176 381 (1) & 34.12 & 19.86 & 0.39 &   [ 1.09,   2.63 ] & 82.06 & 2.41 \\
176 381 (2) & 23.23 & 14.17 & 0.28 &   [ 1.01  ,   2.41 ] & 65.75 & 2.83 \\
176 381 (4) & 18.24 & 11.84 & 0.19 &   [ 0.84,   1.70 ] & 51.57 & 2.83 \\
300 801 (4) & 34.48 & 27.01 & 0.37 &   [ 2.01,   3.80 ] & 118.37 & 3.43 \\
1 071 094 (4)* & 1976.24 & 2279.08 & 2.84 &   [ 7.89  ,   28.05 ] & 5135.51 & 2.60 \\
1 071 094 (8)* & 1627.35 & 1518.88 & 1.13 &   [ 6.17  ,   15.00 ] & 3256.52 & 2.00 \\
1 071 094 (12)* & 1204.97 & 1139.31 & 0.91 &   [ 5.95  ,   12.65 ] & 2471.33 & 2.05 \\
\hline
\end{tabular}
\caption{CPU time for the main computational steps for several problem sizes. The computational work is carried out using parallel computing on multiple processors (1,2,4,8 and 12). Computations are performed on Intel(R) Core(TM) i7-6700HQ CPU \@ 2.60GHz with 16 Go of RAM excepted cases indicated by (*) on Intel(R) Xeon(R) CPU E5-2670 \@ 2.60GHz with 256 Go of RAM. }
\label{tab:time}
\end{center}
\end{table}

\section{Conclusion}
 
A computational method has been developed in order to locate exceptional points (EP) arising in eigenvalue problems depending on a single 
complex-valued parameter. EPs are particular values of the parameter which render the matrix defective. Parametric eigenvalue problems and EPs which are inherent with non-Hermitian systems have attracted considerable attention in classical and quantum mechanics and various numerical techniques have been proposed for their location. 
The aim of the present work is to deal with large size matrices, stemming from discretization methods such as the the finite element method, whereby  the defective eigenvalue is assumed to be of algebraic multiplicity 2. 

The  method is based on the calculation of high order derivatives of a selected pair of eigenvalues, susceptible to coalesce at the EP, using the bordered matrix approach \cite{Andrew:1993}. The loss of regularity of the eigenvalues near EP is circumvented thanks to two functions, which by construction, are regular in the neighbourhood of the EP and which roots allow to identify the location of EPs in the complex plane.

The pair of coalescing  eigenvalues can be recovered as function of the parameter thanks to a truncated Puiseux series expansion which is shown to produce very accurate results  in a certain region of the complex plane  which is limited by the presence of other EPs involving one eigenvalue from the selected pair. Puiseux series provides the topological structure of the eigenvalue Riemann surface. Thus the method allows a precise description of the trajectories of each selected eigenvalues in the complex plane.

For a given pair of eigenvalues susceptible to coalesce, the efficiency
of the method relies on the location of the initial guess. Although there is no a priori on the exceptional point location, the proposed method can be embedded in an adaptive scheme and by exploiting the estimation of the disk of convergence of function $h$, the parametric space can be paved at a reasonable cost.

The method is used in order to investigate the existence of EPs for the analysis of sound wave propagation in guiding structures with absorbing materials. Two examples are chosen in order to illustrate the accuracy and the computational efficiency of the method. In all cases, it is found that the EPs correspond to optimal scenarios whereby the sound attenuation reach a  maximum. 
It is thought that the approach developed in this work could also be particularly relevant in the field of non-Hermitian physics or for other applications dealing with parametric eigenvalue problems like random eigenvalue problems \cite{Ghienne:2017} or stability analysis \cite{Seyranian:2003}.

Because the work presented in this paper is restricted to matrices depending on a single parameter, there is a need for more sophisticated techniques able to handle multiparametric perturbations \cite{Seyranian:2003,Mailybaev:2006} and this could be the subject for future work.

\appendix

\section{Derivatives of the eigenvalues}\label{sec:Deriv0}

The computation of eigenvalues and eigenvectors derivatives  of problem \eqref{eq:PVP}  has received lots of attention during the past decades and several techniques have been developed for this purpose. Appropriate techniques depends on the type of problems, the derivative orders and the number of eigenvalues considered. In \cite{Murthy:1988}, the authors analyse the algorithmic complexity for the treatment of general complex non-Hermitian matrices. 
Methods can be classified into \emph{adjoint}, \emph{direct} and \emph{iterative} \cite{Rudisill:1975,Andrew:1978,Tan:1986} methods, though the latter is not well developed and suited for non-Hermitian matrices \cite{Murthy:1988}.

\emph{Adjoint methods} which use right and left eigenvectors are competitive if only the first derivative of the eigenvalues is needed. The first  derivative of the eigenvector is computed using the knowledge of all eigenvectors. Van der Aa \cite{VanDerAa:2007} extends the technique in order to compute the first derivative for all eigenvalues and the approach has been recently used to speed up the frequency sweep for the computation of waveguide dispersion curves \cite{Krome:2016}. Adjoint methods have also been used for group velocity calculation \cite{Thurston:1977,Moiseyenko:2011}.

\emph{Direct methods}  generally require only right eigenvectors and a single linear system factorization for each eigenvalue regardless of  the number of parameters involved and of the order derivative \cite{Nelson:1976,Andrew:1993}. These methods become attractive from a computational point of view  for high-order derivatives and when the number of selected eigenvalues is small. The global sparse or banded structure of the matrix is conserved and efficient algorithms can be used (see section~\ref{sec:Comput}).

In the present work, we follow the direct method proposed by Andrew et al.~\cite{Andrew:1993} (the reader is referred to the article just mentioned for more details). Eigenvector are normalized using
\begin{equation}\label{eq:norm}
\vv^t \xv(\nuv)=1
\end{equation}
where $\mathbf{v}$ is a constant vector, though other normalization are possible \cite{Andrew:1993}. 
Using the fact that the $n$\textsuperscript{th} order derivative of the eigenvalue problem \eqref{eq:PVP} and of the normalization condition \eqref{eq:norm} should be equal to zero, 
successive derivatives of $(\lambda, \xv)$ can be obtained by the recursive bordered matrix
\begin{equation}
\begin{bmatrix}
\Lv & \partial_\lambda \Lv \xv \\
\vv^t & 0
\end{bmatrix}
\begin{pmatrix}
\xv^{(n)}\\ \lambda^{(n)}
\end{pmatrix}
= \begin{pmatrix}
\Fv_n\\ 0
\end{pmatrix}
\end{equation}
where the right hand side vector $\Fv_n$ contains terms arising from previous order derivatives.  Its general expression is cumbersome because of the nested dependency of $\nu$ in $\Lv (\lambda(\nuv), \nuv)$
though it can be explicitly obtained using Fa\'a Di Bruno formula. When $\Lv$ is a a polynomial function of $\lambda$, which is the case in the examples treated in the present paper, the expression is amenable to a relatively easy treatment.

\bibliographystyle{elsarticle-num}

\end{document}